\documentclass[preprint2]{aastex62}

\newcommand\Msun{\textrm{M}_{\odot}}
\newcommand\Rsun{\textrm{R}_{\odot}}

\usepackage{graphicx}
\usepackage{gensymb}
\usepackage[caption=false]{subfig}
\graphicspath{ {./figures/} }

\received{XXX}
\revised{XXX}
\accepted{XXX}
\submitjournal{ApJ}

\shorttitle{Early SNe~Ia Lightcurves}
\shortauthors{Burke et al.}

\begin{document}

%\title{A Range of Early Lightcurve Behaviors in a Sample of Nine Type Ia Supernovae}
%\title{Distribution of Early Blue Lightcurve Excesses Consistent with SN Ia Progenitor Companions Being Predominantly Nondegenerate}
\title{Early Lightcurves of Type Ia Supernovae are Consistent with Nondegenerate Progenitor Companions}

\correspondingauthor{J. Burke (he, him)}
\email{jburke@lco.global}

\author[0000-0003-0035-6659]{J. Burke}
\affil{Las Cumbres Observatory, 6740 Cortona Dr, Suite 102, Goleta,
CA 93117-5575, USA}
\affil{Department of Physics, University of California, Santa Barbara, CA
93106-9530, USA}

\author[0000-0003-4253-656X]{D. A. Howell}
\affiliation{Las Cumbres Observatory, 6740 Cortona Dr, Suite 102, Goleta, CA 93117-5575, USA}
\affiliation{Department of Physics, University of California, Santa Barbara, CA 93106-9530, USA}

\author[0000-0003-4102-380X]{D. J. Sand}
\affil{Steward Observatory, University of Arizona, 933 North Cherry Avenue, Tucson, AZ 85721-0065, USA}

\author[0000-0002-1546-9763]{R. C. Amaro}
\affil{Steward Observatory, University of Arizona, 933 North Cherry Avenue, Tucson, AZ 85721-0065, USA}

\author[0000-0001-6272-5507]{P. J. Brown}
\affil{Department of Physics and Astronomy, Texas A\&M University, 4242 TAMU, College Station, TX 77843, USA}
\affil{George P. and Cynthia Woods Mitchell Institute for Fundamental Physics \& Astronomy}
 
\author[0000-0003-0123-0062]{J.~E. Andrews}
\affiliation{Gemini Observatory/NSF’s NOIRLab, 670 N. A’ohoku Place, Hilo, Hawai’i, 96720, USA}

\author[0000-0002-4924-444X]{K. A. Bostroem}
\affiliation{Department of Astronomy, University of Washington, 3910 15th Avenue NE, Seattle, WA 98195-0002, USA}
\altaffiliation{DiRAC Fellow}

\author[0000-0001-8818-0795]{Y.~Dong}
\affiliation{Department of Physics and Astronomy, University of California, 1 Shields Avenue, Davis, CA 95616-5270, USA}

\author[0000-0002-6703-805X]{J. Haislip}
\affiliation{Department of Physics and Astronomy, University of North Carolina at Chapel Hill, Chapel Hill, NC 27599, USA}

\author[0000-0002-1125-9187]{D. Hiramatsu}
\affiliation{Center for Astrophysics $|$ Harvard \& Smithsonian, 60 Garden Street, Cambridge, MA 02138-1516, USA}
\affiliation{The NSF AI Institute for Artificial Intelligence and Fundamental Interactions}

\author[0000-0002-0832-2974]{G. Hosseinzadeh}
\affil{Steward Observatory, University of Arizona, 933 North Cherry Avenue, Tucson, AZ 85721-0065, USA}

\author[0000-0003-3642-5484]{V. Kouprianov}
\affiliation{Department of Physics and Astronomy, University of North Carolina at Chapel Hill, Chapel Hill, NC 27599, USA}

\author[0000-0001-9589-3793]{M.~J. Lundquist}
\affiliation{W.~M.~Keck Observatory, 65-1120 M\=amalahoa Highway, Kamuela, HI 96743-8431, USA}

\author[0000-0001-5807-7893]{C. McCully}
\affiliation{Las Cumbres Observatory, 6740 Cortona Dr, Suite 102, Goleta, CA 93117-5575, USA}
\affiliation{Department of Physics, University of California, Santa Barbara, CA 93106-9530, USA}

\author[0000-0002-7472-1279]{C. Pellegrino}
\affiliation{Las Cumbres Observatory, 6740 Cortona Dr, Suite 102, Goleta, CA 93117-5575, USA}
\affiliation{Department of Physics, University of California, Santa Barbara, CA 93106-9530, USA}

\author[0000-0002-5060-3673]{D. Reichart}
\affiliation{Department of Physics and Astronomy, University of North Carolina at Chapel Hill, Chapel Hill, NC 27599, USA}

\author[0000-0003-3433-1492]{L. Tartaglia}
\affiliation{INAF - Osservatorio Astronomico di Padova, Vicolo dell'Osservatorio 5, 35122 Padova, Italy}

\author[0000-0001-8818-0795]{S.~Valenti}
\affiliation{Department of Physics and Astronomy, University of California, 1 Shields Avenue, Davis, CA 95616-5270, USA}

\author[0000-0003-2732-4956]{S.~Wyatt}
\affil{Steward Observatory, University of Arizona, 933 North Cherry Avenue, Tucson, AZ 85721-0065, USA}

\author[0000-0002-2898-6532]{S. Yang}
\affiliation{Department of Astronomy and the Oskar Klein Centre, Stockholm University, AlbaNova, Roslagstullsbacken 21, 114 21 Stockholm, Sweden}

\begin{abstract}
\noindent If Type Ia supernovae (SNe~Ia) result from a white dwarf being ignited by Roche lobe overflow from a nondegenerate companion,
then as the supernova explosion runs into the companion star its ejecta will be shocked,
causing an early blue excess in the lightcurve.
A handful of these excesses have been found in single-object studies,
but inferences about the population of SNe~Ia as a whole have been limited because of the rarity of multiwavelength followup within days of explosion.
Here we present a three-year investigation yielding an unbiased sample of nine nearby ($z<0.01$) SNe~Ia with exemplary early data.
The data are truly multiwavelength,
covering $UBVgri$ and Swift bandpasses,
and also early,
with an average first epoch 16.0 days before maximum light.
Of the nine objects, three show early blue excesses.
We do not find enough statistical evidence
to reject the null hypothesis
that SNe~Ia predominantly arise from Roche-lobe-overflowing single-degenerate systems
($p=0.94$).
When looking at the objects' colors,
we find the objects are almost uniformly near-UV-blue,
in contrast to earlier literature samples which found that only a third of SNe~Ia are near-UV-blue,
and we find a seemingly continuous range of $B-V$ colors in the days after explosion,
again in contrast with earlier claims in the literature.
This study highlights the importance of early, truly multiwavelength, high-cadence data in determining the progenitor systems of SNe~Ia and in revealing their diverse early behavior.
\end{abstract}

\keywords{supernovae}

\section{Introduction} \label{sec:intro}

%Should check out Peter Brown's paper for refs: https://arxiv.org/pdf/1807.10391.pdf

Despite the fact that type Ia supernovae (SNe) were used as standardizable candles to discover the accelerating expansion of the universe and constrain its energy content \citep{riessnobel, perlmutter_99_stretch},
many open questions remain about their progenitor systems. 
The supernovae themselves are understood to be the thermonuclear explosions of carbon/oxygen white dwarfs (WDs) \citep{hoyle},
but the channels by which these explosions occur are poorly constrained.

A large body of literature exists modeling the progenitor systems and explosion mechanisms of SNe~Ia.
The two most commonly invoked progenitor systems are the single-degenerate case \citep{whelan}, where the WD accretes matter slowly from a nondegenerate companion, and the double-degenerate case \citep{iben}, where the source of the extra matter needed to ignite
%ignite might not be the exact word I mean
the WD is a second WD. 
Several varieties of explosion mechanisms have also been modeled,
for instance double-detonation models.
These models have a surface layer of He which detonates,
driving a thermonuclear shock into the WD causing the core to detonate \citep{sim_double_det,polin_double_det}.
Some channels have been suggested which explicitly combine different progenitor systems and explosion mechanisms,
such as the dynamically-driven double-degenerate double-detonation channel \citep{shen6d}.
For reviews, see \citet{howell11}, \citet{wanghan}, \citet{maoz}, and \citet{saurabh_review}.

There are several observational signatures that could distinguish between single- and double-degenerate progenitor systems.
One such signature we focus on in this paper arises in the single-degenerate case:
if the donor star were nondegenerate then the SN ejecta should run into it and get shock-heated.
The shock-heated ejecta would then emit an excess of blue/UV light which could be detected in the SN's early lightcurve,
as was predicted and modeled in \citet{kasen}.
The excess is expected to be only be detectable within $\sim$5 days of explosion --
SNe Ia have a range of rise times,
but this translates to earlier than $\sim$14 days before maximum light \citep{miller_rise_times}.
The strength of this signature is dependent on the companion's size and separation, the velocity of the ejecta, and the viewing angle of the event.
The viewing angle effect alone
means that only approximately 10\% of SNe~Ia arising from this single-degenerate channel would display a strong early excess,
and even then, excellent early, multiwavelength, high-cadence datasets are necessary to identify these features.

Early excesses in SNe~Ia were predicted to arise from this channel before any such features were observed.
A small number of SNe~Ia with early excesses have since been discovered and modeled:
SN~2012cg \citep{marion},
the 02es-like iPTF14atg \citep{cao},
iPTF16abc \citep{miller_16abc},
SN~2017cbv \citep{griffin},
SN~2018oh \citep{wenxiong_18oh, dimitriadis_18oh, shappee_18oh},
the transitional 02es-like SN~2019yvq 
\citep{miller_19yvq, siebert_19yvq, tucker_19yvq, burke_19yvq},
and SN~2021aefx 
\citep{ashall_2021aefx, griffin_2021aefx}.
Other objects have unusual early data,
such as SN~2018aoz,
which showed extreme color evolution in the hours after explosion \citep{ni_18aoz}.
Some papers in the literature have noted objects with weaker early excesses without providing detailed models for any one event \citep{jiang_2018, deckers_ztf_excesses} ,
and other work has established some limits on the rates of early excesses based on non-detections in samples of SNe Ia \citep{bianco_SNLS, brown_2012, olling_2015, fausnaugh_TESS_nondets}.

In addition to companion shocking,
other physical effects can also give rise to early lightcurve features.
An early excess could instead arise from radioactive decay of matter in the outer ejecta,
either from Ni mixed throughout the WD as it burns \citep{piromorozova, magee_maguire}
or from a layer of accreted He on the surface of the WD
\citep[see the double-detonation models described above and in][]{polin_double_det}.
These double-detonation models originally needed large amounts of He 
($\sim$0.2$\Msun$)
in order for the He detonation to disrupt the WD \citep{sim_double_det},
but this amount of He results in spectra and color evolutions which significantly differ from observed SNe~Ia (see Figures 5 and 7 of that paper).
More recent double-detonation models \citep{polin_double_det} require significantly less He 
(down to 0.01$\Msun$),
although the models still have spectra and color evolutions which are atypical for most SNe~Ia.

A complementary line of inquiry relies not on the early photometry,
but on nebular spectroscopy.
In single-degenerate progenitor systems
one would expect the companion star to be stripped to some degree 
and to leave signatures of H in the spectra which become visible in the nebular phase once the SN has faded 
\citep{marietta_nebula_halpha,botyanski_halpha,dessart_nebular_H}.
This signature has not been observed
for $>$100 SNe~Ia with nebular spectra,
often to very constraining limits of $M_{\rm{stripped \ H}}<10^{-3} \Msun$ \citep[for recent compilations see][]{maguire_nebular,dave_nebular_halpha,tucker_nebular_halpha}. 
It has been observed in a handful of cases, namely
SN~2016jae \citep{16jae_nebular_halpha},
SN~2018cqj \citep{18cqj_nebular_halpha},
and
SN~2018fhw \citep{18fhw_nebular_halpha}.
Curiously,
all three SNe are underluminous,
with $M_{B,\rm{peak}}$ fainter than $-18$,
and no H$\alpha$ has been observed for any objects with an early UV excess
including for SN~2017cbv \citep{dave_17cbv_nebular},
SN~2018oh \citep{dimitriadis_18oh_nebular,tucker_18oh_nebular},
or SN~2021aefx \citep{ashall_2021aefx, griffin_2021aefx}.
In addition to this paper focusing on early-time photometry,
a companion paper (Sand et al. 2022, in prep.) will focus on the nebular spectra of objects in this sample,
although no signatures of H$\alpha$ are detected.

X-ray non-detection limits can also be used to probe the circumstellar environments of SN Ia progenitors.
Non-detections throughout the literature are collected in \citet{17cbv_xray},
which also presents stringent new limits for two objects:
SNe 2017cbv (discussed here) and 2020nlb.
The observations of these two objects are among the strongest X-ray limits for any SNe Ia,
and effectively rule out a symbiotic giant star companion for their progenitor systems
(see Figure 6 of that paper).

A sample of early SNe~Ia lightcurves needs early, high-cadence, multiwavelength photometry to characterize any potential early excesses.
Such a sample has become easier to build up in recent years due to the proliferation of time-domain surveys such as
ASAS-SN \citep{shappee_asassn, kochanek_asassn},
ATLAS \citep{tonry_atlas},
DLT40 \citep{tartaglia_dlt40},
%Gaia \citep{gaia},
%Kepler \citep{kepler},
%TESS \citep{tess},
and ZTF \citep{bellm_ztf}.
Combining discoveries from these surveys with followup from facilities like
Las Cumbres Observatory \citep[LCO;][]{lco} and the 
Neil Gehrels Swift Observatory \citep[{Swift};][]{Gehrels04}
allows for characterization of SNe~Ia at early times across a wide range of UV, optical, and NIR wavelengths at daily or sub-day cadences.
%These surveys and space missions are detecting more and more SNe,
%and can provide data at younger and younger phases.

We present an analysis of the early lightcurves of nine nearby SNe~Ia,
observed by LCO and Swift,
and in most cases discovered by DLT40,
a sub-day cadence SN survey of nearby galaxies.
In total we have 6,110 data points,
beginning an average of 16.0 days before maximum light.
Utilizing LCO allows for sub-day cadence observations from $U$-band through $i$-band:
this high cadence multiwavelength followup is the ideal way to characterize early SNe Ia as it can probe observational signatures at short timescales and across the optical spectrum,
even into the near-UV
where companion interaction signatures are expected to dominate at early times.
We search for signatures of companion interaction in our sample and find three such cases,
including in the data of SN~2018yu, published here for the first time.
We compare to separate model grids in an effort to distinguish between different progenitor systems and explosion mechanisms,
although ultimately we favor companion interaction models.
We also measure rise times and colors to compare to literature values.

This paper is organized as follows.
In Section \ref{sec:observations} we detail the different data sources and reduction methods.
In Section \ref{sec:analysis} we analyze the lightcurves,
measuring standard parameters such as stretch and peak magnitudes.
Using those parameters
we fit our data with different models and characterize early excesses
in Section \ref{sec:model},
and in Section \ref{sec:color} we compare the color evolution of our sample to samples of SNe~Ia from the literature.
We discuss the pros and cons of our adopted companion interaction models in Section \ref{sec:discussion},
before concluding in Section \ref{sec:conclusions}.
%All code used during analysis and plot generation is available on github\footnote{\url{https://github.com/jfrostburke/earlysneia}}.

\begin{figure*}[htbp]
\begin{center}
\includegraphics[width=0.985\textwidth]{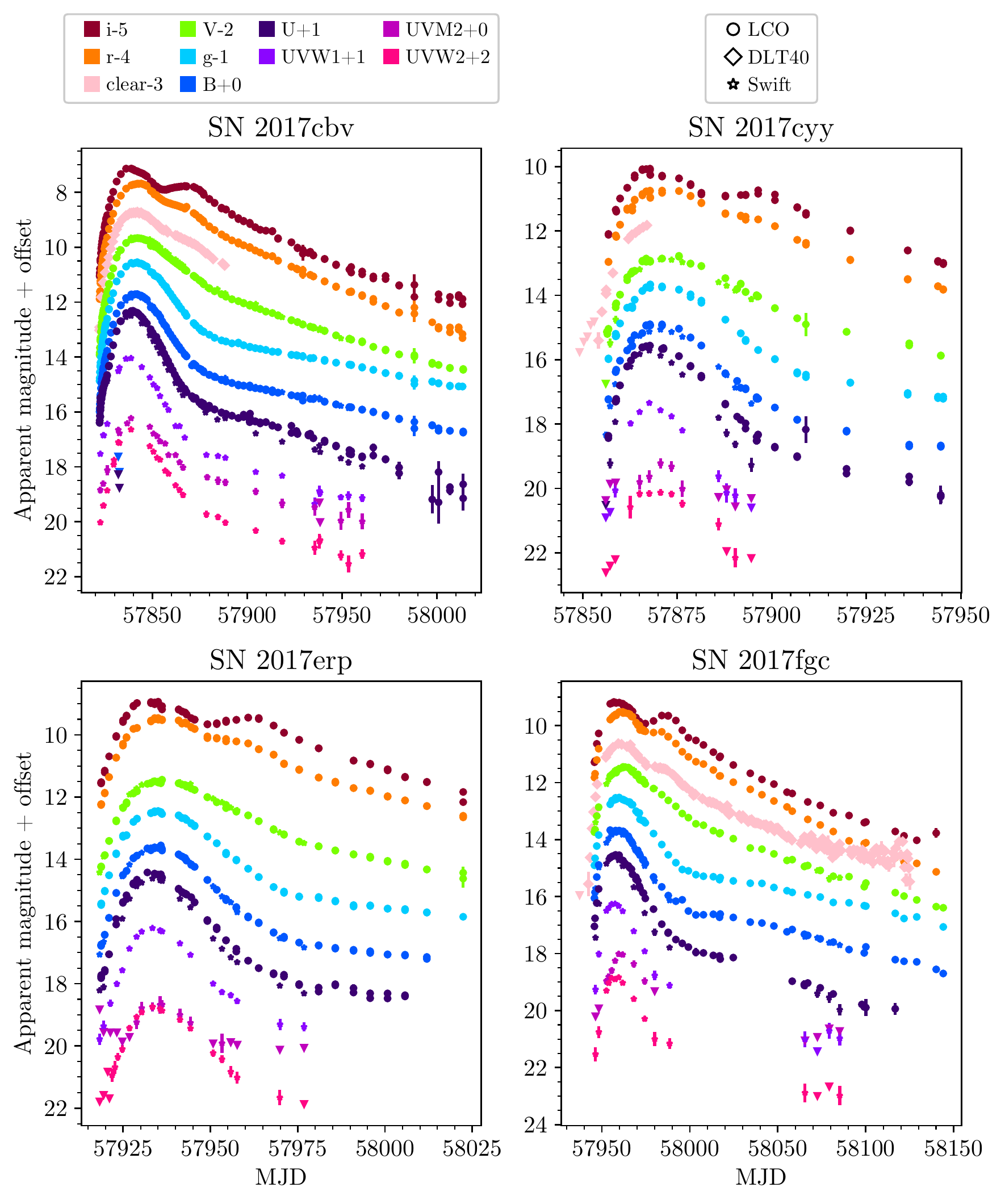}
\caption{
Full datasets (DLT40+LCO+Swift) for all objects in the sample.
Lower limits are included as downward-pointing triangles.
}
\label{fig:lightcurves}
\end{center}
\end{figure*}

\begin{figure*}[htbp]
\ContinuedFloat
\begin{center}
\includegraphics[width=0.985\textwidth]{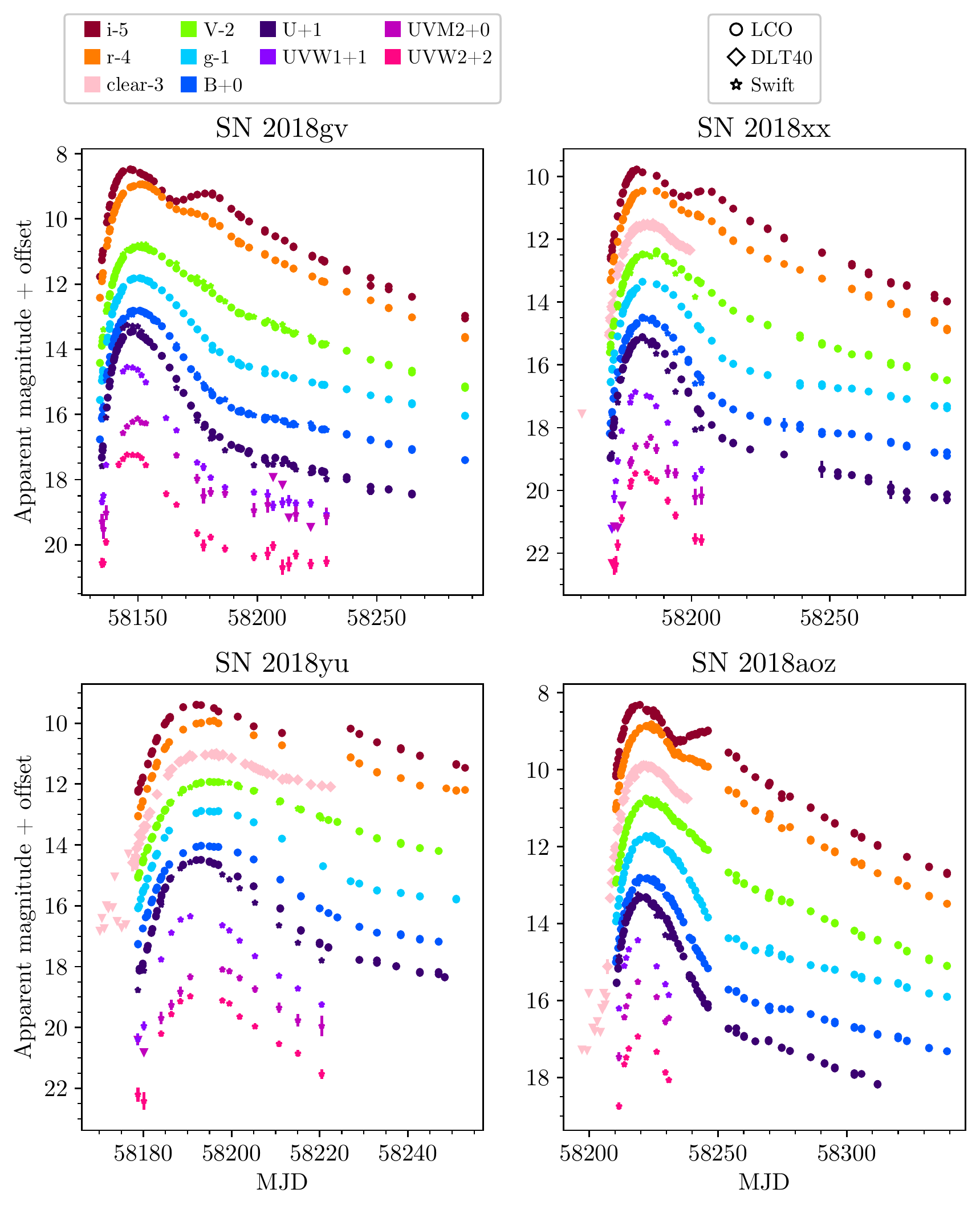}
\caption{
Continued.
}
\end{center}
\end{figure*}

\begin{figure}[htp]
\ContinuedFloat
\begin{center}
\includegraphics[width=0.47\textwidth]{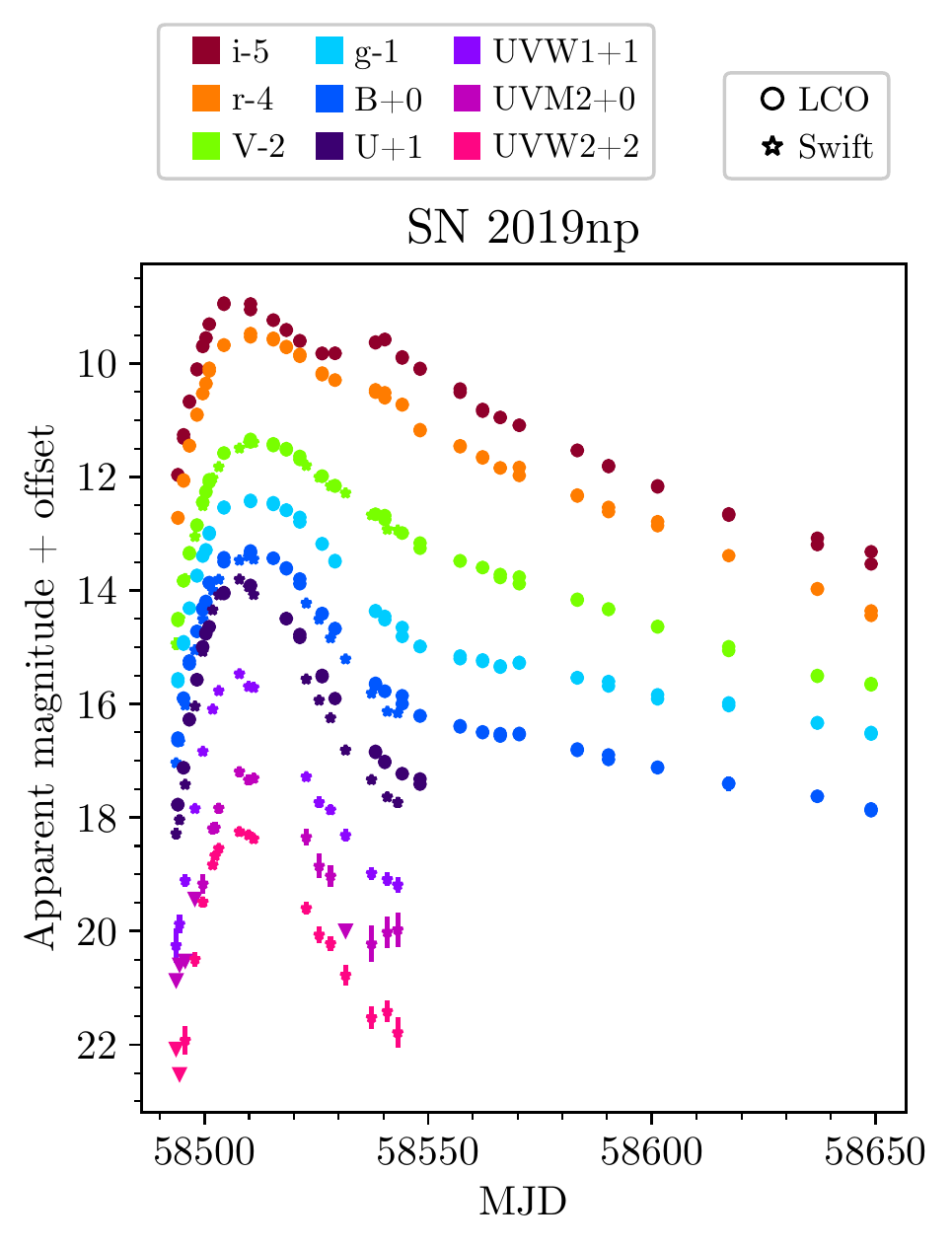}
\caption{
Continued.
}
\end{center}
\end{figure}

\section{Observations and Data Reduction} \label{sec:observations}

Of the nine objects in this sample,
most were discovered by the DLT40 survey
and all have high-cadence multiwavelength lightcurves from LCO,
with additional UV photometry from Swift.
Our full dataset is presented in Figure \ref{fig:lightcurves}.

Although a focus of this paper is the modeling of early excesses (see Section \ref{ssec:blue_bumps}),
we did not select the sample based on objects with early excesses.
We constructed this 
sample attempting to be as unbiased as possible, using the following criteria:
we consider all nearby ($z<0.01$) spectroscopically normal SNe~Ia discovered between 2017 and 2019 (inclusive),
which were thoroughly observed ($>$300 images) by the Global Supernova Project at LCO.
These criteria yield a sample of 11 objects,
and we impose two additional requirements:
that the objects be discovered within a few days of explosion 
(this removes one object, SN~2017gah)
and that the objects do not have excessive reddening 
($A_{V}<1 \Rightarrow E(B-V)<0.322$),
since accurate absolute magnitudes are needed to judge the strength or absence of any early excesses 
and since high extinction decreases the signal-to-noise in the crucial early blue data,
making it more difficult to detect early blue bumps
(this removes another object, SN~2017drh, which has $E(B-V)>1$).
These criteria result in a sample of nine SNe~Ia.
The sample is not perfectly unbiased
(e.g. the Global Supernova Project only elects to follow a subset of nearby SNe Ia given finite observing time),
but it represents the state of the art for samples of nearby SNe Ia with early, high-cadence, truly multiwavelength data.
The sample should not be biased to skew towards objects with early bumps.

\subsection{Discovery}

The majority of SNe presented here (SN~2017cbv, SN~2017cyy, SN~2017fgc, SN~2018aoz, SN~2018xx, SN~2018yu) were discovered by the Distance Less Than 40 Mpc survey \citep[DLT40;][]{tartaglia_dlt40}.  DLT40 is a $\sim$12-24 hr cadence search for nearby SNe, targeting galaxies within $D$$\lesssim$40 Mpc, and designed with the goal of discovering $\sim$5--10 SNe per year within a day of first light.  Initially DLT40 operated a single 0.4-m telescope at Cerro Tololo Inter-American Observatory (CTIO) in Chile, with a second telescope at Meckering Observatory (Western Australia) joining the search in 2017 December.  Further technical details of this program are discussed elsewhere \citep{Yang19,Bostroem20}, and we utilize the full DLT40 lightcurves and detection limits here (seen as the pink points in Figure \ref{fig:lightcurves}) to help provide useful constraints on the explosion epoch and rise times.

DLT40 photometry is taken in a ``clear" (Meckering Observatory) or ``open" (CTIO) filter,
and is calibrated to \textit{r}-band photometry of the AAVSO Photometry All-Sky Survey
\citep[APASS;][]{apass}.  For the purposes of this work, we will refer to this DLT40 light curve data as ``clear'' for convenience.

The three objects in this sample not discovered by DLT40 
(i.e. SNe 2017erp, 2018gv, and 2019np)
were all discovered by the amateur astronomer Koichi Itagaki
\citep{17erp_discovery, 18gv_tns, 19np_discovery}.

\subsection{Followup}

After discovery,
we performed intensive high-cadence multiwavelength followup with the telescopes of LCO
in addition to UV--optical observations from Swift.
Photometric reduction for the LCO images was accomplished using \texttt{lcogtsnpipe} \citep{valenti16},
a PyRAF-based pipeline.
All SNe were bright enough and far enough separated from their host galaxies that their reductions did not require template image subtraction.

We report all \textit{UBV} observations in Vega magnitudes.
We calculate zeropoints for Landolt filters using same-night same-telescope observations of Landolt standard fields \citep{stetson}. 
We report all \textit{gri} observations in AB magnitudes.
We calculate zeropoints for Sloan filters using using Sloan catalog stars in the fields of each SN where possible \citep{sdss}.
If there are no Sloan catalog stars in the field,
we used APASS catalog stars for zeropoint calibration \citep{apass}.

Additionally, all SNe presented here have also been observed by
Swift using the the Ultra-Violet Optical Telescope \citep[UVOT;][]{Roming05}.
Observations were reduced using the pipeline associated with the Swift Optical Ultraviolet Supernovae Archive \citep[SOUSA;][]{Brown_etal_2014_SOUSA} and the zeropoints of \citet{Breeveld10}.

Our full dataset is presented in Figure \ref{fig:lightcurves}.

\section{Analysis} \label{sec:analysis}

In this section we will discuss analysis done on the photometry to derive standard lightcurve parameters.
Results are summarized in Table \ref{table:lc_params}.

\begin{deluxetable*}{ccccccc}
\tablecaption{Lightcurve Parameters}
\tablehead{\colhead{SN} & \colhead{$\textrm{MJD}_{\rm{max}}$} & \colhead{$m_{B}$} & \colhead{DM} & \colhead{$M_{B}$} & \colhead{$\Delta m_{15}$} & \colhead{$s$}}
\startdata
2017cbv & $57840.29 \pm 0.09$ & $11.715 \pm 0.008$ & $30.657 \pm 0.007$ & $-19.55 \pm 0.01$ & $0.953 \pm 0.009$ & $1.031^{+0.002}_{-0.002}$ \\
2017cyy & $57870.12 \pm 0.04$ & $14.89 \pm 0.01$ & $33.443 \pm 0.009$ & $-19.46 \pm 0.03$ & $1.067 \pm 0.007$ & $0.927^{+0.007}_{-0.007}$ \\
2017erp & $57934.41 \pm 0.04$ & $13.60 \pm 0.02$ & $32.32 \pm 0.02$ & $-19.48 \pm 0.03$ & $0.987 \pm 0.008$ & $0.939^{+0.007}_{-0.006}$ \\
2017fgc & $57959.47 \pm 0.06$ & $13.70 \pm 0.02$ & $32.06 \pm 0.02$ & $-19.46 \pm 0.04$ & $1.061 \pm 0.005$ & $1.08^{+0.02}_{-0.02}$ \\
2018gv & $58149.64 \pm 0.01$ & $12.847 \pm 0.006$ & $32.125 \pm 0.006$ & $-19.48 \pm 0.01$ & $1.014 \pm 0.005$ & $1.0381^{+0.0003}_{-0.0003}$ \\
2018xx & $58183.90 \pm 0.02$ & $14.58 \pm 0.02$ & $33.39 \pm 0.02$ & $-19.43 \pm 0.03$ & $1.159 \pm 0.003$ & $0.847^{+0.002}_{-0.002}$ \\
2018yu & $58194.32 \pm 0.04$ & $14.002 \pm 0.007$ & $33.011 \pm 0.007$ & $-19.51 \pm 0.01$ & $0.952 \pm 0.005$ & $1.043^{+0.004}_{-0.003}$ \\
2018aoz & $58222.18 \pm 0.02$ & $12.846 \pm 0.004$ & $31.623 \pm 0.004$ & $-19.12 \pm 0.01$ & $1.432 \pm 0.002$ & $0.881^{+0.002}_{-0.002}$ \\
2019np & $58509.64 \pm 0.06$ & $13.37 \pm 0.03$ & $32.80 \pm 0.03$ & $-19.65 \pm 0.06$ & $0.98 \pm 0.01$ & $1.05^{+0.03}_{-0.03}$
\enddata
\tablecomments{
Lightcurve parameters for objects in the sample,
using the methodology detailed in Section \ref{sec:analysis}.
We measure the peak magnitude in $B$-band ($m_B$) by fitting to the observed lightcurve (i.e. not dereddened),
whereas we have dereddened the peak absolute magnitude ($M_B$).
The $\Delta m_{15}$ is derived from \texttt{SNooPy},
and does not directly correspond to $\Delta m_{15}(B)$.
$\textrm{MJD}_{\rm{max}}$ is also specifically for $B$-band data.
Objects are consistent with zero host extinction,
besides 
SN~2017fgc ($E(B-V) = 0.231 \pm 0.006$),
SN~2017erp ($E(B-V) = 0.098 \pm 0.006$),
and
SN~2018xx ($E(B-V) = 0.037 \pm 0.005$).
}
\label{table:lc_params}
\end{deluxetable*}

\begin{deluxetable}{cccc}
\tablehead{\colhead{SN} & \colhead{$z$} & \colhead{Host name} & \colhead{Host type}}
\startdata
2017cbv & 0.003999 & NGC5643 & SABcd \\
2017cyy & 0.009777 & ESO091-015 & SABm \\
2017erp & 0.006174 & NGC5861 & SABc \\
2017fgc & 0.007722 & NGC474 & SA0 \\
2018gv & 0.00527 & NGC2525 & SBc \\
2018xx & 0.00999 & NGC4767 & E \\
2018yu & 0.008112 & NGC1888 & SBc pec \\
2018aoz & 0.005801 & NGC3923 & E4-5 \\
2019np & 0.00452 & NGC3254 & SABc
\enddata
\caption{
Host information for objects in the sample.
}
\label{table:sn_hosts}
\end{deluxetable}

\subsection{\texttt{SNooPy}}\label{ssec:snoopy}

We make use of the \texttt{SNooPy} package \citep{snoopy} to measure several lightcurve parameters,
including the host extinction and distance modulus.
Following \citet{burke_19yvq}
we do exclude both $U$-band data 
(since \texttt{SNooPy} cannot fit that data)
% it cannot fit the data at all, it does not have U templates
and $i$-band data
\citep[since the variation in secondary IR maxima leads to overfitting that feature, with worse overall fits; see][for an investigation of the variation in secondary $i$-band maxima and its effect on SN Ia parameter estimation]{pessi_i_band}.

We do fits using the default \texttt{EBV\_model} and the \texttt{fitMCMC()} procedure,
enforcing $R_{V,\rm{host}} = 3.1$.
Multiple objects
(SN~2017cbv,
SN~2017cyy,
SN~2018yu,
SN~2018aoz)
converge to small negative (and therefore unphysical) values of $E(B-V)$.
This can also be seen when comparing the MW-dereddened $B-V$ colors to the ``Lira Law"
\citep[see Equation 1 and Figure 1 of][]{Phillips99},
as those objects are slightly bluer than the expected Lira Law template.
As such, when doing the MCMC fit we impose a uniform prior on \texttt{EBVhost} ranging from 0 to 1.
We visually inspect fits to ensure that they are reasonable.

Most objects converge to 0 host extinction,
as expected from the fact that they are well-separated from their host galaxies.
The two highest-reddening objects are
SN~2017erp ($E(B-V)_{\rm{host}}= 0.098 \pm 0.006$)
and
SN~2017fgc ($E(B-V)_{\rm{host}}= 0.231 \pm 0.006$).
SN~2017erp was the subject of study in \citet{brown_2017erp},
which found two possible fits to the host extinction:
$E(B-V)=0.10$ $(R_V = 3.1)$, or 
$E(B-V)=0.18$ $(R_V = 1.9)$.
Our value is consistent with the $R_V=3.1$ value in that paper.
SN~2017fgc has also been studied before,
and our value is consistent with that found in
\citet{zeng_17fgc} ($E(B-V)=0.17 \pm 0.07$).

We convert each object's $E(B-V)$ to per-filter extinction values in two different ways,
depending on the filter.
For $UBVgri$ data,
we use the \citet{schlafly_finkbeiner} recalibration of the \citet{schlegel_dust_map} dust maps,
accessed by the Python package \texttt{extinction} \citep{extinction_zenodo}.
For Swift filters we use the method described in \citet{peter_brown_swift_extinction},
specifically the red-corrected coefficients listed in Table 1 of that paper.

We also use \texttt{SNooPy} to measure the distance modulus for each object.
Host information for the objects in the sample is listed in Table \ref{table:sn_hosts}.
Eight of the nine objects are in NGC galaxies,
each of which has a variety of redshift-independent distance measurements cataloged on the NASA/IPAC Extragalactic Database (NED\footnote{\url{https://ned.ipac.caltech.edu/}}).
The measurements for a single object can have a wide range:
the average standard deviation for host galaxies' distance moduli is 0.71 mags,
with measurements for one object (NGC 1888) ranging from 27.98 \citep{1888_distance_lo} to 33.06 \citep{1888_distance_hi}.
In an effort to use a uniform methodology for measuring distance
we adopt the values inferred by \texttt{SNooPy},
which are consistent with the range of values listed on NED for each host.

Lastly, we measure the peak brightness in $B$-band using the \texttt{get\_max} method in \texttt{SNooPy}.
The absolute magnitudes range from $-19.12\pm0.01$ for the faintest object (SN 2018aoz)
to $-19.65\pm0.06$ for the brightest (SN 2019np) --
this distribution conforms to the range expected for normal SNe Ia \citep[see e.g. Figure 7 of][where the majority of their sample has $-20<M_{B}<-19$]{ashall_2016}.

The time of maximum (for $B$-band),
peak apparent magnitude,
and dereddened peak absolute magnitude,
in addition to the distance modulus and $\Delta m_{15}$,
are all reported in Table \ref{table:lc_params}.

\subsection{Stretch}\label{ssec:stretch}

Stretch ($s$) is a single parameter useful in modeling SNe~Ia
\citep{perlmutter_97_stretch,perlmutter_99_stretch,goldhaber_stretch},
measured by taking a flux-normalized SN Ia lightcurve and stretching it temporally to maximize overlap with a template.
Although relations exist to convert $\Delta m_{15}$ to $s$ \citep[e.g.][]{ganeshalingam_rise_times},
we independently measure $s$ for the objects in our sample.

We make use of the MCMC framework of the \texttt{lightcurve\_fitting} package
\citep{griffin_lightcurvefitting},
which utilizes the \texttt{emcee} package \citep{emcee}.
We create a simplified version of the \texttt{CompanionShocking} model that eliminates the actual companion shocking calculation,
but keeps the stretch-correction to an $s=1$ SiFTO template \citep{sifto}.

We limit the data to around peak (-10 to +30 days),
and we additionally limit to only $B$-band data as done in \citet{goldhaber_stretch}.
We monitor MCMC walkers to ensure successful burn-in and convergence.
Stretches are reported in Table \ref{table:lc_params}.

The measured stretch values also confirm that the sample consists of normal SNe~Ia:
the average stretch of the sample is $1.03 \pm 0.04$ 
(consistent with the expected value of 1 for samples of normal SNe~Ia).
The object most discrepant from $s=1$ is SN~2018xx ($s=0.847 \pm 0.002$),
still above the cutoff of $s>0.8$ typically used for differentiating fast-declining/subluminous SNe~Ia \citep{gonzalez_gaitan_2011_stretch}.

\section{Model Fits and Early Excesses} \label{sec:model}

In this section we use the parameters measured in the above section,
notably the distance modulus and the host extinction from which we derive absolute magnitudes,
to compare to a variety of early SN Ia models.

\subsection{The Search for Blue Bumps}\label{ssec:blue_bumps}

\subsubsection{Description of companion shocking models}\label{sssec:companion_models}

We fit the early LCO+DLT40 data of our objects
(to $t_{\rm{max}}+5$ days)
with models similar to those described in 
\citet{griffin},
\citet{burke_19yvq},
and \citet{griffin_2021aefx},
making use of the \texttt{lightcurve\_fitting} package \citep{griffin_lightcurvefitting}.
The models consist of two components:
a template lightcurve representing a $s=1$ SN Ia from \texttt{SiFTO}
(as described in Section \ref{ssec:stretch}),
to which is added excess flux arising from a companion shocking interaction,
which can dominate at early times.

For the \texttt{SiFTO} template,
we refer to \citet{sifto} for the full details of its construction.
But for the earliest epochs,
the per-filter templates are extrapolated to zero flux with a power law of the form
$f=a(t-t_{\rm{exp}})^{2} + b(t-t_{\rm{exp}})^{3}$,
where $t_{\rm{exp}}$ is the date of zero flux.
This approximately matches the so-called ``expanding fireball" model of early SN Ia flux from \citet{arnett_82},
which predicted a parabolic rise,
and which was experimentally verified by the earliest data of SN 2011fe \citep{nugent_11fe}.
Our \texttt{SiFTO} templates only cover $UBVgri$ data, so we cannot use Swift data bluer than $U$.
However, these near-UV data are often severely overpredicted by our analytic model with its grey opacity and lack of UV-reprocessing in the ejecta \citep[see e.g.][]{griffin, griffin_2021aefx},
so we would most likely exclude those data even if we could them.

The companion shocking component is used in the form of the corrected analytic approximations from \citet{kasen} (Equations 22 and 25) included in \citet{griffin} (Equations 1 and 2),
reproduced here:

\begin{figure*}[t!]
\begin{center}
\includegraphics[width=0.98\textwidth]{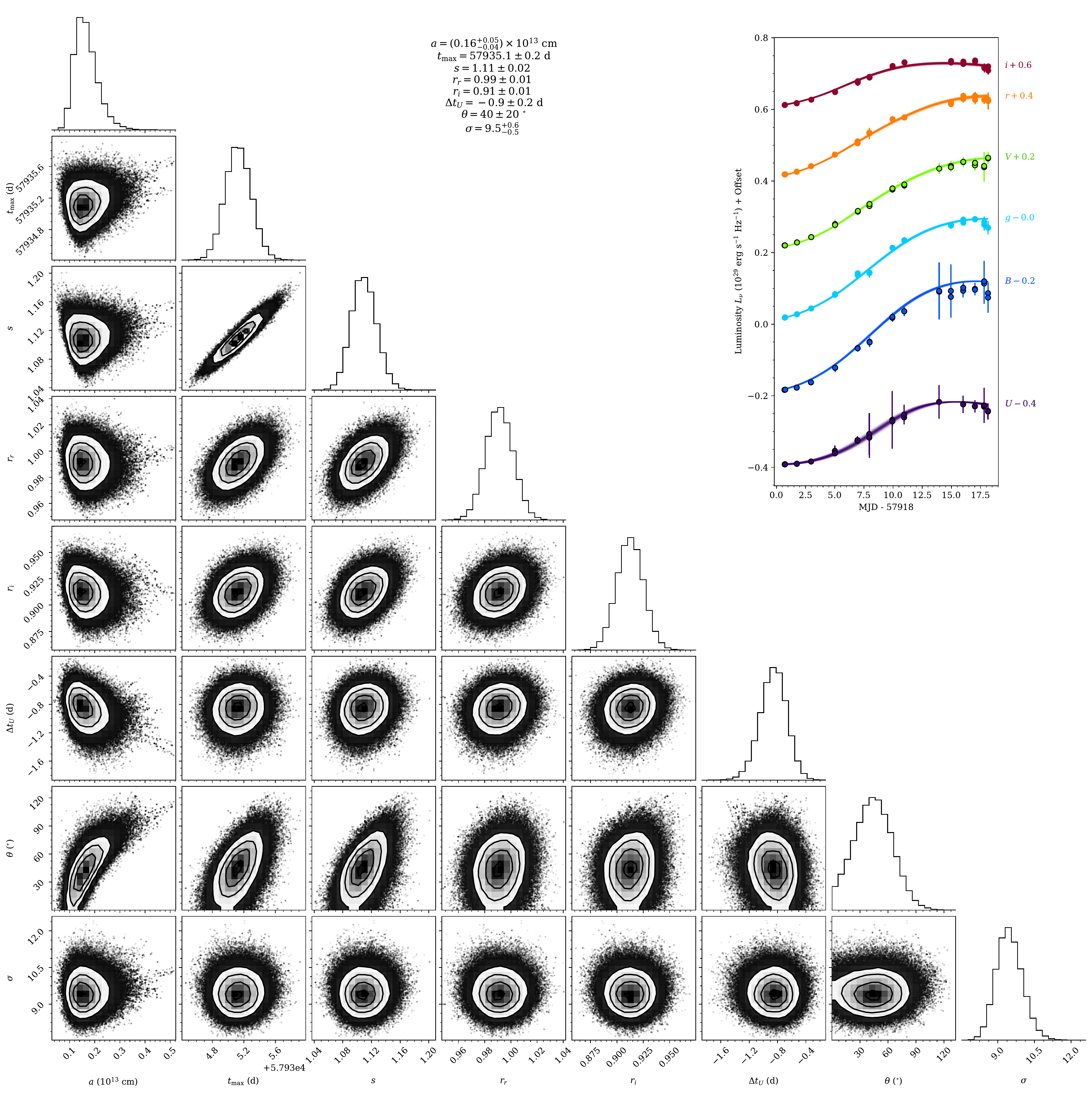}
\caption{
Corner plot of the model described in Section \ref{sssec:companion_models} for SN~2017erp.
Errors in the inset lightcurve have been multiplied by the best-fit value of the $\sigma$ parameter.
}
\label{fig:17erp_corner}
\end{center}
\end{figure*}

\begin{equation}
    R_{\rm{phot}} = (2700\textrm{ }\Rsun) x^{1/9} \kappa^{1/9} t^{7/9}
\end{equation}
\begin{equation}\label{eqn:kasen_T}
    T_{eff} = (25000\textrm{ K}) a^{1/4} x^{1/144} \kappa^{-35/144} t^{-37/72}
\end{equation}
where
$a$ is the binary separation (in units of $10^{13} \textrm{ cm} \approx 144\textrm{ }\Rsun$),
$\kappa$ is the opacity (in units of the electron scattering opacity, we set $\kappa=1$),
$t$ is the time since explosions (in days),
and $x$ is a quantity consisting of the ejecta mass and speed:
\begin{equation}
    x \equiv \left( \frac{M}{\textrm{M}_{\rm{Ch}}} \right) \left( \frac{v}{10000 \textrm{ km s}^{-1}}  \right)^{7}
\end{equation}
where $M$ is the ejecta mass,
$\textrm{M}_{\rm{Ch}}$ is the Chandrasekhar mass,
and $v$ represents a transition velocity in the original model's ejecta between inner (high-density) and outer (low-density) ejecta.

The above radius and temperature are combined to form a blackbody luminosity component representing the companion-shocked ejecta.
As \citet{kasen} showed,
this luminosity is highly dependent on viewing angle
due to the inherently asymmetric conditions of the ejecta--companion collision.
Following the semi-analytic form of \citet{brown_2012},
we include a multiplicative factor on the luminosity of the companion interaction component of the form

\begin{equation}\label{eqn:kasen_factor}
    f = \left( 0.5 \cos \theta + 0.5 \right) \times \left( 0.14 \theta^{2} - 0.4 \theta + 1 \right)
\end{equation}
where $\theta$ is the viewing angle in radians.
This factor is bound between 1 
($\theta=0$, perfectly aligned observer and shock)
and 0
($\theta=\pi$, perfectly misaligned).

Because the $x$ parameter influences the temperature extremely weakly
(i.e. primarily only affects the luminosity of the shock component, and therefore is highly covariant with viewing angle)
and because the relevant velocity is not an easily measured physical value,
we set $x$ to its 
fiducial value of 1
\citep[in order-of-magnitude agreement with the best-fit value in e.g.][]{griffin}.
This results in a total of 8 parameters:

\begin{enumerate}
    \item $a$, the companion separation of the shock component
    \item $t_{\rm{max}}$, the time of $B$-band maximum light for the stretch component
    \item $s$, the stretch applied to the stretch component
    \item $r_{r}$, a factor on the $r$-band flux of the stretch component
    \item $r_{i}$, a factor on the $i$-band flux of the stretch component
    \item $\Delta t_{U}$, a shift on the time of $U$-band maximum for the stretch component
    \item $\theta$, the viewing angle (which determines a multiplicative factor on the shock component as shown in Equation \ref{eqn:kasen_factor})
    \item $\sigma$, a multiplicative factor on the errors of the data to account for error underestimation.
\end{enumerate}
As described in \citet{griffin},
the $r_{r}$ and $r_{i}$ factors are included because the combination of luminosity decrease and temperature decrease of the shock component results in a non-negligible amount of shock flux in redder filters even out to peak brightness,
even though the primary signature of companion interaction is an early UV excess.
As in \citet{griffin_2021aefx},
which used similar models to fit the data of SN 2021aefx (another SN Ia with an early excess),
we find that fits are improved by temporally shifting the $U$-band component of the stretch template,
which could reflect the fact that SNe Ia are less homogeneous in the UV \citep[see e.g.][]{milne_uv_colors}.
The $\Delta t_{U}$ parameter is reported as the shift (in days) applied to the $U$-band component.

Additionally,
the model implicitly measures the rise times of the objects.
Since the rise time of the template 
(from time of first light to $B$-band maximum)
is 17.19 days,
the rise time of an object fit by this model is simply that value multiplied by its measured stretch.

\begin{figure*}[ht!]
\begin{center}
\includegraphics[width=0.47\textwidth]{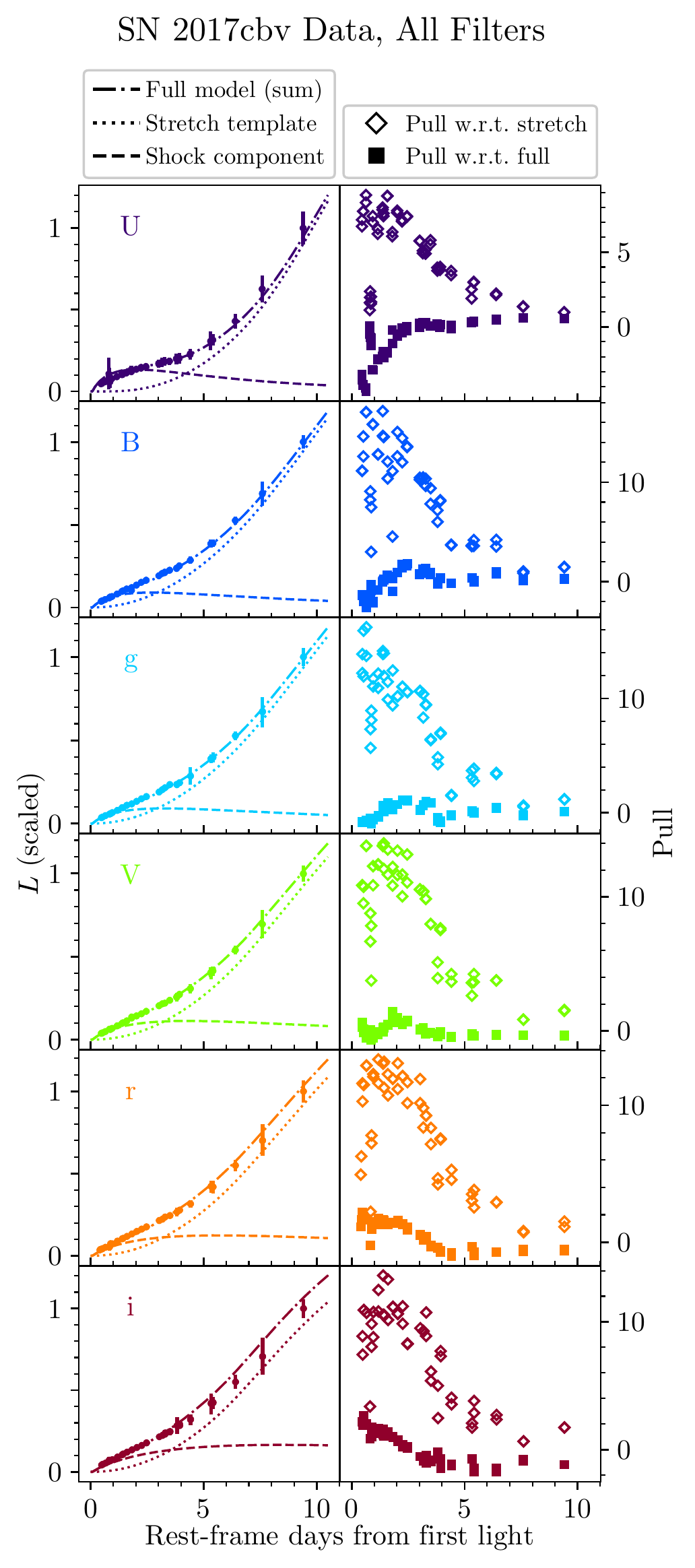}
\includegraphics[width=0.47\textwidth]{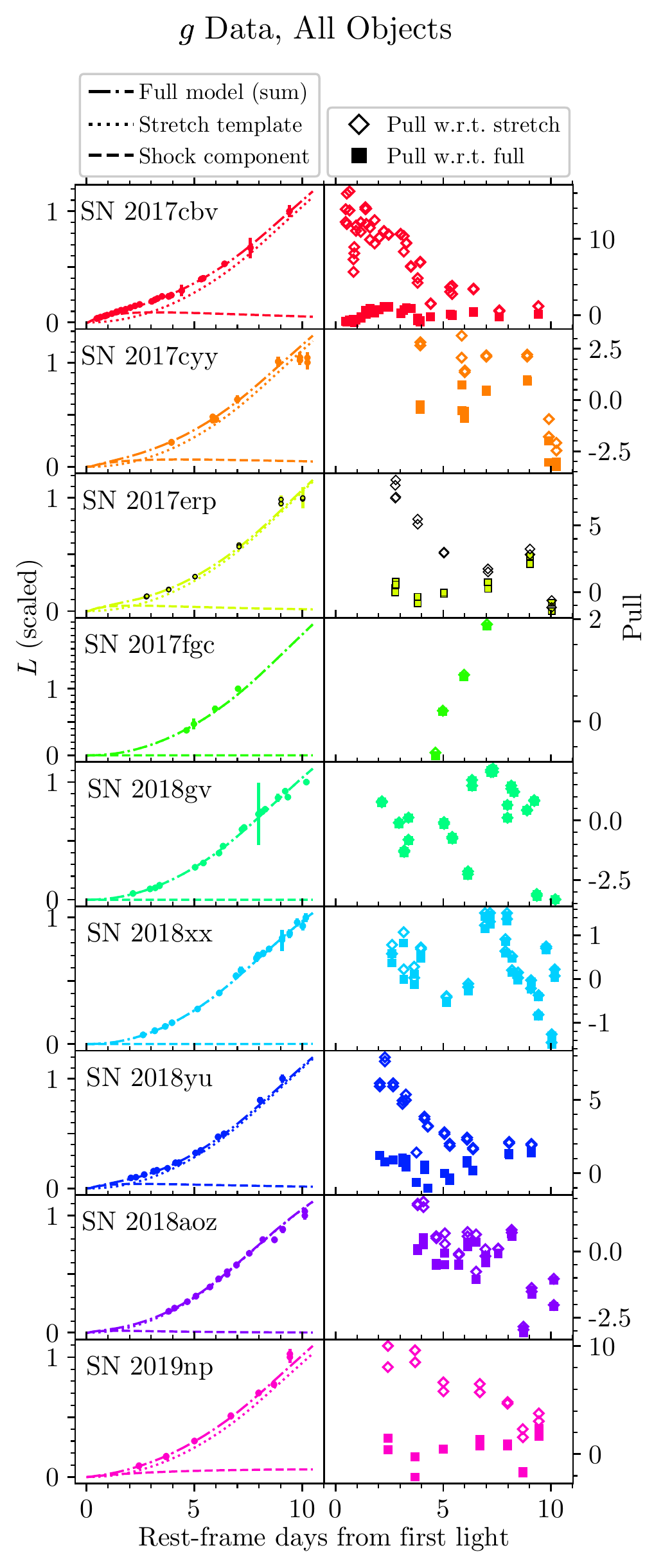}
\caption{
Fits and residuals for the models described in Section \ref{sssec:companion_models}
The left panel shows all filters for one object, SN~2017cbv.
The residuals are plotted both with respect to the full model (squares)
and with respect to just the stretched template (diamonds),
to visualize the strength of the early excess.
Residuals are plotted as the normalized ``pull,"
defined as $(L_{\rm{data}}-L_{\rm{model}})/(\sigma_{\rm{data}}\times\sigma)$
(i.e. number of corrected standard deviations away from the model,
with $\sigma$ being the best-fit value in Table \ref{table:kasen_parameters}}).
The right panel shows the same style of plot but looking in one filter ($g$) across all objects.
The variety of early behavior is apparent in the strengths and durations of the discrepancy between the two sets of pulls.
\label{fig:kasen_residuals}
\end{center}
\end{figure*}

%Same as the previous figure,
%but instead of looking at all filters for one object,
%looking at all objects in one filter ($g$).
%The variety of early behavior is apparent in the strengths and durations of the discrepancy between the two sets of pulls.

Different versions of this model have been used to fit objects with obvious early excesses
\citetext{\citealp[e.g. SN~2017cbv in][]{griffin};
\citealp[SN~2018oh in][]{dimitriadis_18oh};
\citealp[SN~2019yvq in][]{burke_19yvq}
\citealp[and][]{miller_19yvq};
\citealp[SN~2021aefx in][]{griffin_2021aefx}},
but including the viewing angle as a parameter allows us to self-consistently fit all objects,
regardless of whether or not they have an early excess
(since those without an excess will converge to high viewing angles).
Figure \ref{fig:17erp_corner} shows the corner plot of the fit for SN~2017erp,
an object that has been previously noticed to have a weak early excess \citep{jiang_2018}.
As expected,
due to the weak excess the model converges to a slightly off-axis shock
($\theta = 40 \pm 20 \degree$).

\subsubsection{Best-fit parameters}\label{sssec:kasen_best_fits}

\begin{deluxetable*}{ccccccccc}
\tablehead{\colhead{SN} & \colhead{$a\textrm{ (R}_{\odot}\textrm{)}$} & \colhead{$t_{\textrm{max}}$ (MJD)} & \colhead{$s$} & \colhead{$r_{r}$} & \colhead{$r_{i}$} & \colhead{$\Delta t_{U}$ (days)} & \colhead{$\theta$ $(\degree)$} & \colhead{$\sigma$}}
\startdata
\textbf{2017cbv} & $55.2^{+4.7}_{-4.2}$ & $57840.32^{+0.10}_{-0.10}$ & $1.077^{+0.006}_{-0.006}$ & $0.952^{+0.011}_{-0.011}$ & $0.866^{+0.012}_{-0.011}$ & $-0.83^{+0.08}_{-0.08}$ & $34^{+5}_{-5}$ & $12.57^{+0.44}_{-0.42}$ \\
2017cyy & $98.2^{+67}_{-38}$ & $57870.99^{+0.20}_{-0.22}$ & $1.044^{+0.021}_{-0.026}$ & $1.029^{+0.019}_{-0.019}$ & $0.857^{+0.024}_{-0.025}$ & $-0.37^{+0.21}_{-0.22}$ & $94^{+24}_{-29}$ & $8.92^{+0.54}_{-0.48}$ \\
\textbf{2017erp} & $23.1^{+7.3}_{-5.4}$ & $57935.13^{+0.17}_{-0.16}$ & $1.109^{+0.018}_{-0.017}$ & $0.991^{+0.010}_{-0.010}$ & $0.912^{+0.013}_{-0.013}$ & $-0.85^{+0.18}_{-0.19}$ & $44^{+20}_{-21}$ & $9.45^{+0.56}_{-0.50}$ \\
2017fgc & $390^{+650}_{-350}$ & $57959.81^{+0.21}_{-0.21}$ & $1.078^{+0.017}_{-0.017}$ & $0.983^{+0.013}_{-0.014}$ & $0.966^{+0.015}_{-0.015}$ & $0.35^{+0.09}_{-0.09}$ & $176^{+3}_{-8}$ & $3.45^{+0.30}_{-0.26}$ \\
2018gv & $400^{+680}_{-390}$ & $58149.13^{+0.05}_{-0.05}$ & $0.995^{+0.004}_{-0.004}$ & $1.008^{+0.006}_{-0.006}$ & $0.979^{+0.009}_{-0.009}$ & $-0.26^{+0.03}_{-0.03}$ & $177^{+2}_{-8}$ & $7.00^{+0.29}_{-0.26}$ \\
2018xx & $140^{+690}_{-130}$ & $58183.78^{+0.06}_{-0.06}$ & $0.904^{+0.005}_{-0.007}$ & $1.074^{+0.008}_{-0.007}$ & $0.949^{+0.008}_{-0.009}$ & $-0.32^{+0.06}_{-0.06}$ & $171^{+7}_{-47}$ & $5.19^{+0.24}_{-0.23}$ \\
\textbf{2018yu} & $28.9^{+8.6}_{-6.9}$ & $58194.20^{+0.12}_{-0.11}$ & $1.008^{+0.012}_{-0.012}$ & $0.939^{+0.008}_{-0.008}$ & $0.885^{+0.008}_{-0.008}$ & $-0.78^{+0.05}_{-0.05}$ & $59^{+14}_{-15}$ & $4.36^{+0.21}_{-0.20}$ \\
2018aoz & $8.18^{+4.7}_{-3.2}$ & $58221.76^{+0.06}_{-0.06}$ & $0.878^{+0.007}_{-0.007}$ & $0.959^{+0.006}_{-0.006}$ & $0.962^{+0.008}_{-0.008}$ & $-0.05^{+0.04}_{-0.04}$ & $67^{+25}_{-30}$ & $3.52^{+0.16}_{-0.15}$ \\
2019np & $890^{+370}_{-410}$ & $58510.18^{+0.14}_{-0.16}$ & $1.073^{+0.011}_{-0.015}$ & $0.933^{+0.011}_{-0.012}$ & $0.882^{+0.013}_{-0.014}$ & $-1.32^{+0.24}_{-0.25}$ & $150^{+4}_{-7}$ & $5.78^{+0.42}_{-0.39}$
\enddata
\caption{
Best-fit parameters for the models described in Section \ref{ssec:blue_bumps}.
We have bolded object names where we confidently identify an early excess arising from a companion shocking interaction.
Due to the nature of the models,
the rise times of these objects are directly related to the stretch: $t_{\rm{rise}}=s \times 17.19$ days 
(the rise time of the template).
As noted in the text,
the companion separations 
of objects with no detectable excesses (i.e. high viewing angles)
are extremely poorly constrained and essentially just reflect our priors,
so their best-fit values should not be taken as physical truth.
}
\label{table:kasen_parameters}
\end{deluxetable*}

Table \ref{table:kasen_parameters} lists the best-fit parameters for all objects in the sample.
We identify objects as having an early excess based on three criteria:
if we have data within five days of first light,
the typical epochs for early excesses in \citet{kasen}
(this criterion is necessary but does not exclude any objects in this sample);
if the best-fit viewing angle is less than 90$\degree$;
and if the $g$ residuals with respect to the stretch template 
(see Figure \ref{fig:kasen_residuals})
show a clear systematic,
decreasing from $>5\sigma$ at the earliest epochs to being consistent with the stretch template $>$10 days after explosion.
We choose $g$ for our main comparison because systematics in the residuals arising from a simplified temperature evolution (discussed in more detail soon) are minimized in this band,
allowing the cleanest comparison with the model.
We identify three objects (in bold) which have an early excess:
SN~2017cbv \citep[excess modeled in][]{griffin},
SN~2017erp \citep[excess noted in][but modeled here for the first time]{jiang_2018},
and SN~2018yu (excess modeled here for the first time).
The other objects converge to viewing angles which are too high to confidently model the progenitor system
(e.g. SNe 2017fgc and 2018gv, which have $\theta>170\degree$, resulting in uncertainties on the companion separation of $>$100$\Rsun$),
or have residuals which do not show the systematic behavior expected from a fading shock component (SNe 2017cyy and 2018aoz).

SN~2018aoz is also the subject of study in \citet{ni_18aoz},
where data earlier than presented here show an early red bump,
as opposed to the early blue bump expected from these models.
\citet{ni_18aoz} attribute this behavior to absorption from Fe-group elements in the outer ejecta,
further proof that extremely early data 
(in this case within 12.4 hr of first light)
can reveal key clues to the progenitor systems of SNe~Ia.
SN~2019np was also modeled in \citet{2019np_sai},
which noted unusual early behavior and attributed it to Ni mixing.
The residuals of SN~2019np do show behavior partly consistent with other early-bump objects,
starting off with a $>5\sigma$ discrepancy with respect to the stretch template at the earliest epochs and decreasing over time,
but the timescale of this behavior is significantly longer than for the other three objects with early bumps,
lasting $\sim$10 days instead of the more expected $\sim$5 days.
The best-fit parameters of this object are unusual,
with a highly off-axis explosion coupled with the largest best-fit companion separation in the sample.
Due to its unusual place in parameter space we do not confidently claim it as having an early bump arising from companion interaction.

Figure \ref{fig:kasen_residuals} show the fits and residuals of these models.
As stated earlier,
the models consist of a stretched template
(dotted lines)
and the shock component
(dashed lines).
These are added together to the full model
(dash-dotted lines).
The left panels show the model and its components compared to the data.
The right panels show the residuals of the model,
both with respect to the full model (squares)
and with respect to just the stretch component (diamonds).
Ideally
the squares should cluster around 0 and the diamonds should show the strength of any deviation from the stretched template.

The left half of Figure \ref{fig:kasen_residuals} shows the fits for a single object
(SN~2017cbv, which has the clearest early excess)
across all available filters.
In most filters the early lightcurve shows a clear deviation of more than $10\sigma$ from the stretch template.
Even though the $\Delta t_{U}$ parameter is small (average value of $-0.33$ days across the sample),
the fits show why it is necessary:
without the temporal shift the stretch template overpredicts the data at most epochs even without an additional shock component.
This is possibly explained by the fact that
SNe~Ia show more diversity in the UV than they do in the optical
\citep{ellis_08_uv_diversity,milne_uv_colors, foley_uv_diversity, brown_uv_colors},
making the construction of an accurate and universal $U$-band $s=1$ template more difficult.
In fact no $s=1$ template exists for the Swift filters bluer than \textit{U},
which is why they are excluded from these fits.
However, even including $\Delta t_{U}$ the full model still displays systematics at the earliest times:
the red bands start off overpredicted and slope to zero pull,
while the bluer bands start off underpredicted.
This is indicative of inaccurate temperature evolution:
immediately post-explosion the predicted temperature is too high,
resulting in too much flux in the bluer bands.
This can be explained by the fact that the models are an analytic approximation: note that in Equation \ref{eqn:kasen_T},
$T \propto t^{-37/72}$, i.e. $T(t=0)=\infty$.
This is clearly unphysical and results in the inaccurate temperature effects seen above,
but still represents the start of the art for SN Ia companion shocking models more than a decade after the publication of \citet{kasen}, which is the origin of these models.
As discussed in \citet{griffin} these models make several simplifications,
such as representing the shock flux as a pure blackbody component,
with the spherically symmetric ejecta having a grey, non-time-varying opacity and a simplified density profile.
We attempt to capture the clearly non-spherically-symmetric nature of the explosion with the viewing angle parameter from \citet{brown_2012},
although its semi-analytic formulation is a simplification as well.
A more realistic model which includes the line-blanketing ejecta reprocessing the shock flux is needed to resolve the systematics which are present in Figure \ref{fig:kasen_residuals} and which will continue to be present in the future as more SNe Ia with early excesses are discovered.
We strongly encourage the development of such models
even though they will be more computationally expensive than the purely analytic formulation here.

\begin{figure}[t!]
\begin{center}
\includegraphics[width=0.47\textwidth]{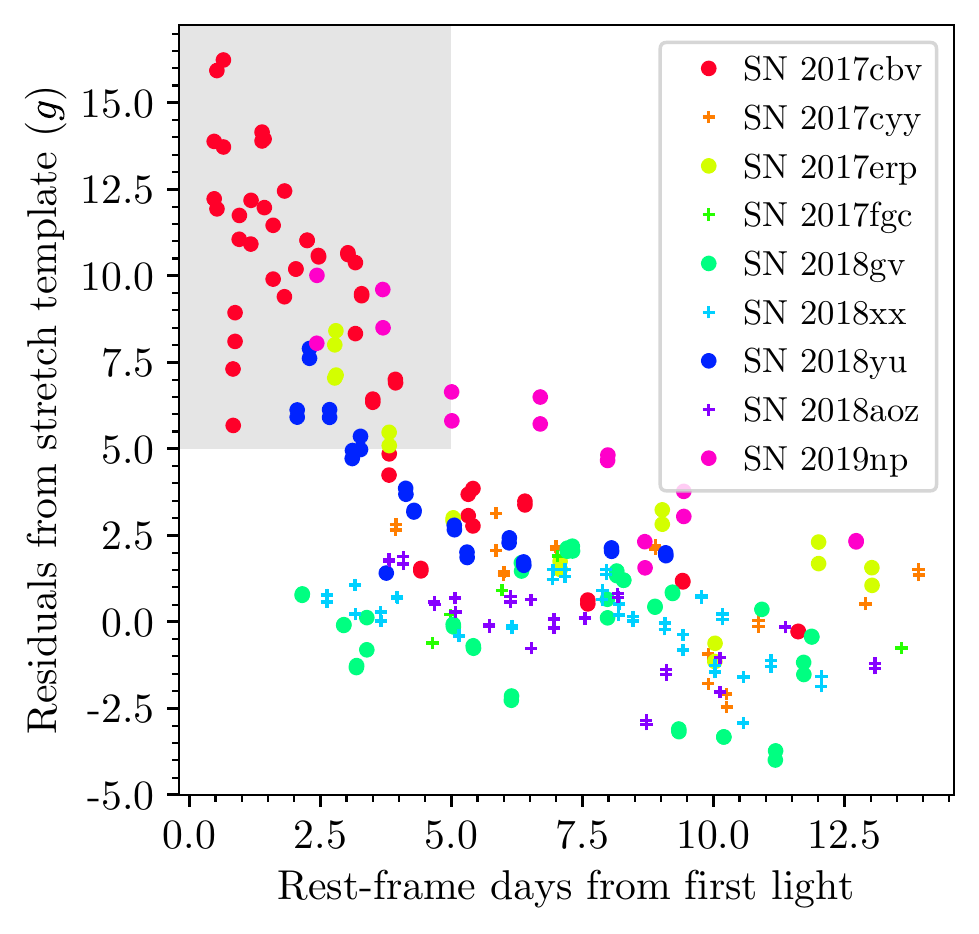}
\caption{
The same residuals 
(with respect to the stretch template, $g$-band)
as in Figure \ref{fig:kasen_residuals},
overplotted together to more easily compare between different objects.
We identify objects with residuals in the grey box 
(earlier than 5 days from first light, stronger than $5\sigma$)
as having an early excess,
except for SN~2019np for reasons discussed in the text.
}
\label{fig:kasen_residuals_all}
\end{center}
\end{figure}

The right half of Figure \ref{fig:kasen_residuals} shows the fits and residuals for all objects in a single filter 
($g$).
The residual plots show a variety of behavior,
from SN~2017cbv's $15\sigma$ discrepancy between model components to SN~2017fgc showing effectively no excess at all.
The residuals with respect to the stretch template for all objects are also plotted in Figure \ref{fig:kasen_residuals_all},
for easier comparison.

\begin{figure}[t!]
\begin{center}
\includegraphics[width=0.47\textwidth]{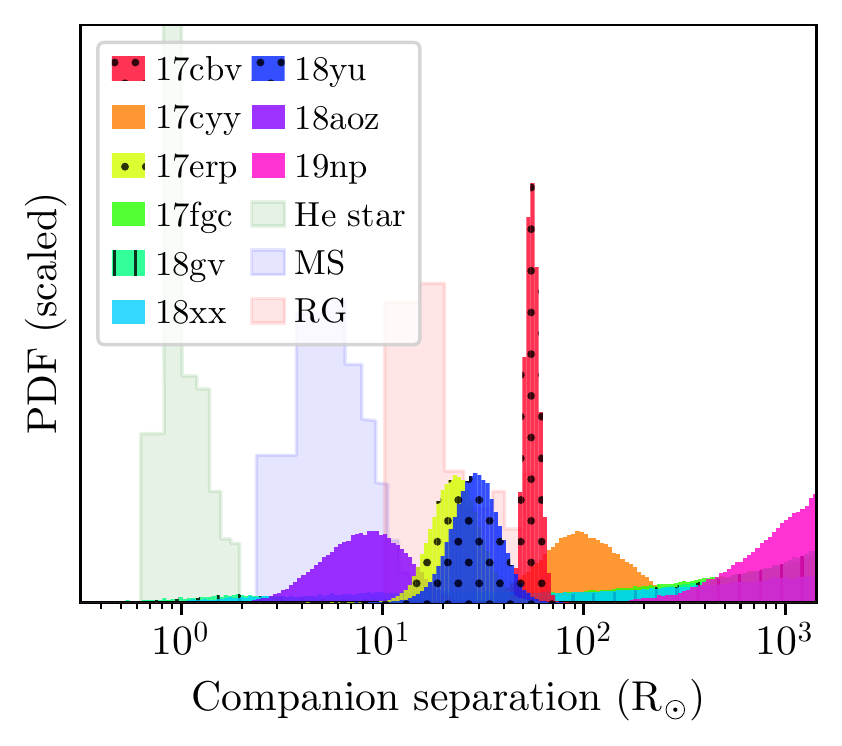}
\includegraphics[width=0.47\textwidth]{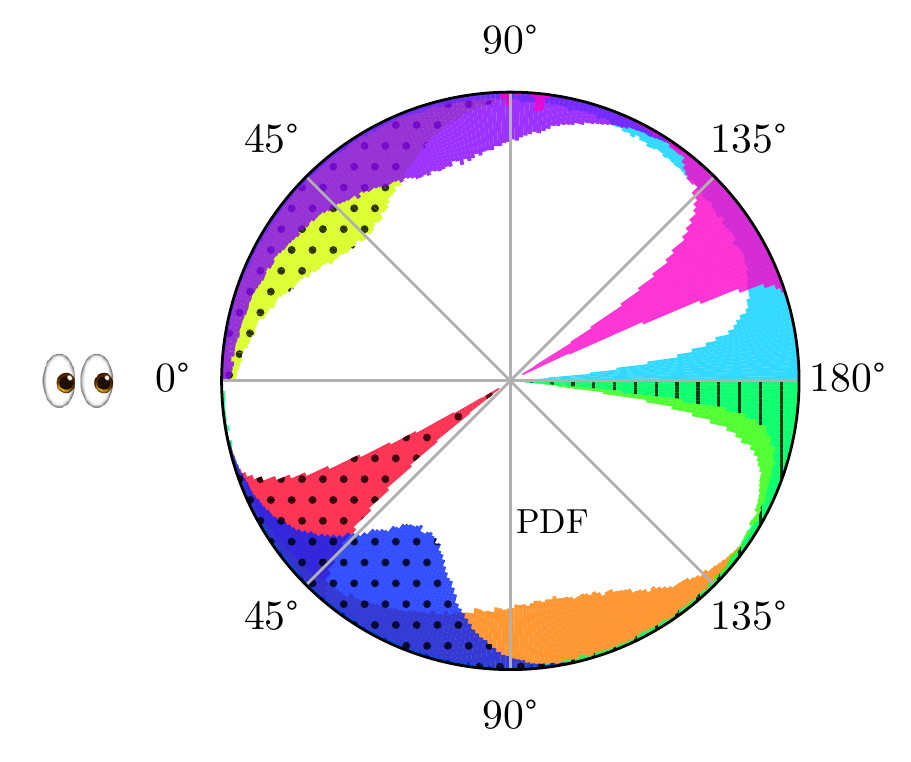}
\caption{
Posteriors for the $a$ and $\theta$ parameters from the fits described in Section \ref{sssec:kasen_best_fits}.
% comment about how we can have a numbers at all addressed in the text now
The three objects with early excesses 
(SNe 2017cbv, 2017erp, and 2018yu)
are emphasized with a dotted hatch.
SN 2018gv also has a hatch, to distinguish itself from SN 2017fgc since the two have such similar posteriors.
We have included predictions for the companion separations from different classes of donor stars 
(He star, main sequence, or red giant)
from \citet{liu_2015}.
The top and bottom half of the viewing angle plot are equivalent,
and objects are placed in different halves solely to minimize overlaps between posteriors.}
\label{fig:kasen_a_theta_posteriors}
\end{center}
\end{figure}

For one last point of comparison with the best-fit parameters,
the companion separations of SN Ia progenitor systems can be inferred for different types of systems,
as was done in \citet{liu_2015}.
That paper used binary evolution models to estimate companion separations for single-degenerate SN Ia progenitor systems in cases where the donor star is a He star,
a main sequence (MS) star,
or a red giant (RG).
We compare their predictions with our best-fit companion separations in Figure \ref{fig:kasen_a_theta_posteriors},
where we plot the posterior probability distributions for that parameter in the top panel.
The maximum predicted separation in \citet{liu_2015} is $55.8\ \Rsun$,
consistent with the best-fit value for SN 2017cbv ($55.2^{+4.7}_{-4.2}\ \Rsun$).
The difficulty in accurately modeling binary stellar evolution makes 
us question whether the upper limit in \citet{liu_2015} is a strict limit:
\citet{kasen} used companion separations of 
29 $\Rsun$ ($2 \times 10^{12} \rm{cm}$) for a 6 $\Msun$ main sequence companion \citep[a companion separation firmly in the red giant distribution of][]{liu_2015}
and a separation of
290 $\Rsun$ ($2 \times 10^{13} \rm{cm}$) for a red giant companion (well outside the red giant range plotted in Figure \ref{fig:kasen_a_theta_posteriors}).
The best-fit companion separations for the objects with early excesses are consistent with red giant companions \citep[according to][]{liu_2015},
although SNe 2017erp and 2018yu do have values similar to the separation for a main sequence companion used in \citet{kasen}.
For the objects which have converged to extremely off-axis systems,
the posterior distributions for their companion separations are essentially flat in linear space.
This means the prior has essentially not been modified,
and the ``best-fit" values listed in Table \ref{table:kasen_parameters} only represent our priors and not physical parameters of the systems.
The posteriors only appear peaked to high separation in Figure \ref{fig:kasen_a_theta_posteriors} because the bins are logarithmic to more easily distinguish objects.

We also include the posteriors for the viewing angle in the bottom panel of Figure \ref{fig:kasen_a_theta_posteriors}.
The plot is oriented to physically replicate the viewing angles with respect to the observer (represented by the eyes).
Objects with no detectable excesses have converged to high viewing angles,
with multiple objects having a maximum likelihood at $\theta=180\degree$.

\begin{figure}[t!]
\begin{center}
\includegraphics[width=0.47\textwidth]{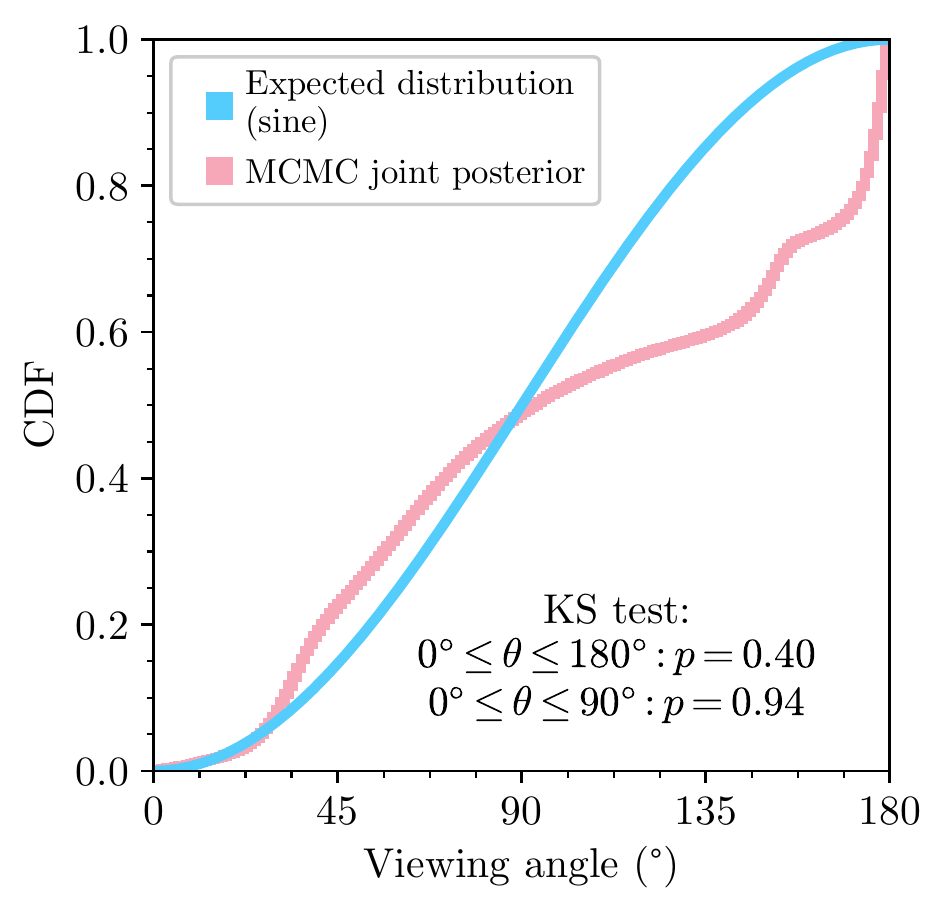}
\caption{
The CDF of viewing angle posteriors compared with the expected distribution.
There is not enough evidence to reject the null hypothesis that SNe Ia predominantly arise in single-degenerate systems.
}
\label{fig:theta_CDF}
\end{center}
\end{figure}

\begin{figure}[t!]
\begin{center}
\includegraphics[width=0.47\textwidth]{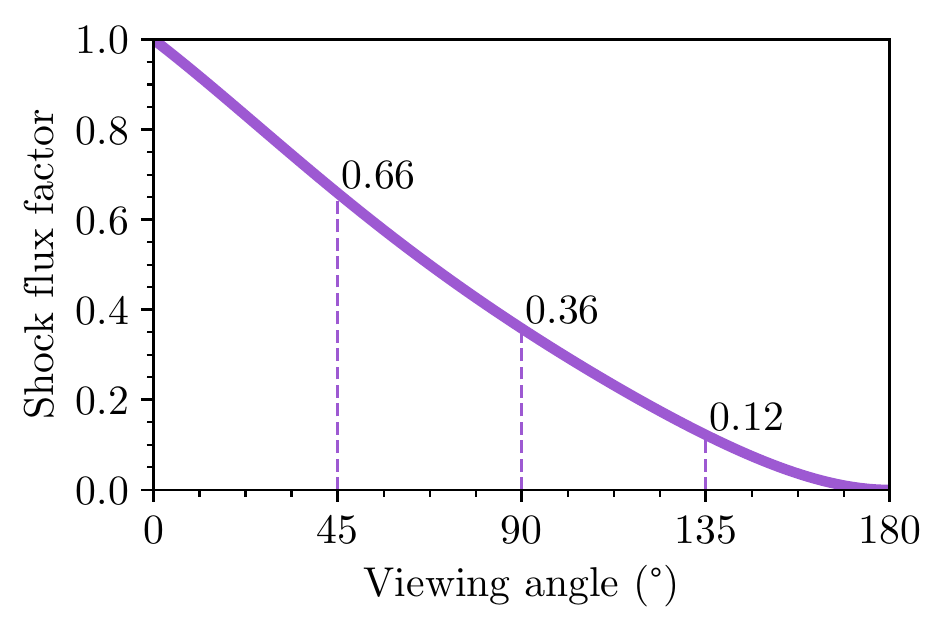}
\caption{
The multiplicative factor on the companion interaction shock flux,
i.e. Equation \ref{eqn:kasen_factor},
originally from \citet{brown_2012}.
Shock flux is extremely diminished for $\theta > 135 \degree$,
which could explain why multiple objects converge to $\theta=180\degree$ when no obvious excess is detected
(see Figure \ref{fig:kasen_a_theta_posteriors}),
leading to the largest discrepancy between the expected and measured distributions
(see Figure \ref{fig:theta_CDF}).
}
\label{fig:theta_function}
\end{center}
\end{figure}

The cumulative distribution function (CDF) is plotted in Figure \ref{fig:theta_CDF}
and compared with the expected distribution (a sine function) if all SNe~Ia arose from Roche-lobe-overflowing single-degenerate systems.
Comparing the two CDFs with a Kolmogorov-Smirnov test yields a $p$-value of 0.40.
However, as is clear in Figures \ref{fig:kasen_a_theta_posteriors} and \ref{fig:theta_CDF},
objects with no detectable early excess have typically converged to a totally off-axis explosion,
with multiple objects having $\theta=180\degree$ as the maximum likelihood.
The reason for this is clear when looking at Figure \ref{fig:theta_function},
which simply plots Equation \ref{eqn:kasen_factor} 
(the multiplicative factor on the shock flux as a function of viewing angle).
The function is non-linear,
making it extremely hard to distinguish between highly off-axis shocks:
at $\theta=135\degree$ the signal from the shock will be below the detection limit, and will be functionally indistinguishable from a perfectly misaligned system.
Therefore we report two $p$-values on Figure \ref{fig:theta_CDF}:
the formal $p$-value across the whole distribution,
and also the $p$-value calculated using $0\degree<\theta<90\degree$, 
i.e. the viewing angle range which we can reasonably hope to distinguish given the quality of our data.
Using this second $p$-value 
(i.e. using the value calculated from parameter ranges which we believe are meaningful since they result in detectable and distinguishable signals),
the two distributions are consistent at the
$p=0.94$ level.
We find no statistical evidence for a preferred viewing angle, which might indicate a contribution from a different model for the early excess.
There is not enough evidence to reject the null hypothesis,
which is that SNe Ia predominantly arise in single-degenerate systems.
This claim will be tested more robustly with larger samples of SNe~Ia.

Despite the fact that we only find three early excesses in our sample,
we fit all objects with the same set of models for the sake of consistency.
Across the literature there is a wide variety of methodologies used in fitting the rising lightcurves of SNe~Ia,
which we will address in a future work 
(Burke et al. 2022, in prep.).

\subsubsection{Early excess rate}\label{sssec:early_excess_rate}

Three of the objects in our sample display an early excess.
Assuming simple Poissonian errors,
this corresponds to an early excess rate of $33 \pm 19 \%$.
Our data for all objects in this sample are early enough that we believe we would recover excesses for any object that had one
which lasted on the timescale of $\approx$5 days,
as expected from \citet{kasen}.

Our value is consistent with other estimates for the rate of early excesses in SNe~Ia,
such as 
\citet{deckers_ztf_excesses} (early excess rate of $18 \pm 11 \%$)
and
\citet{magee_models} (rate of $\sim$22\%, 5 out of 23 SNe),
even though these studies do not adopt the same null hypothesis as we do 
(single-degenerate systems with companion interaction).
\citet{jiang_2018} find a rate of 100\% in 91T-like/99aa-like SNe~Ia,
based on six out of six overluminous SNe~Ia with early excesses.
Due to a lack of 91T-like SNe~Ia in our sample we cannot confirm this claim,
although out of the four brightest objects in our sample three of them have an early excess.
It is possible that early excesses preferentially occur in bright SNe~Ia,
but larger samples including 91T-like SNe~Ia are needed to thoroughly test that claim.

\subsubsection{Rise times}\label{sssec:rise_times}

The models described above can also be used to measure rise times for the objects in our sample.
We extrapolate the models to the time of first light,
defined as the time when photons can first diffuse out of the SN ejecta 
\citep[i.e. when the SN is first visible, which is not necessarily identical to when it exploded, see][]{piro_dark_phase}.
The time of first light is directly related to the inferred stretch:
since the stretch template has a rise time of 17.19 days,
the rise time is simply $t_{\rm{rise}}=s\times17.19$ days.
Combining this with the $t_{\rm{max}}$ parameter results in the MJD of first light.

The average rise time of our sample is 
$16.81 \pm 1.26$ days.
Stretch-correcting this
(i.e. using the stretch for the full lightcurve,
listed in Table \ref{table:lc_params}, not the stretch inferred from the rising lightcurve)
yields a consistent value with smaller variance:
$17.17 \pm 0.86$ days.
These values are consistent with some recent measurements from the literature:
\citet{gonzalez_gaitan_12} find an average rise time of
$17.02^{+0.18}_{-0.28}$ days 
and
\citet{papdogiannakis_19} find an average of 
$16.8^{+0.6}_{-0.5}$ days.
However these values are mildly inconsistent with other reported average rise times,
such as those from
\citet{firth_rise_times} (average of $18.98 \pm 0.54$ days)
or
\citet{miller_rise_times} (average of $18.9$ days,
although with a range for individual obejcts from 15 to 22 days).

We again note a wide range of methodologies used to measure the rise times of SNe~Ia,
which we will discuss in more detail in a later work (Burke et al. 2022, in prep.).

\subsubsection{Early spectra}\label{sssec:early_spec}

As a final point of comparison for these models,
early spectra can be used to test their predictions.
The models directly predict that,
for SNe Ia with early excesses,
their early flux should be dominated by a blackbody shock component.
Although it is beyond the scope of this paper,
this prediction has been tested for some objects with early excesses,
where it does appear that their early spectra are diluted with a blackbody component.
So far this has been rigorously tested for
SN 2012cg \citep{marion} and SN 2019yvq \citep{burke_19yvq},
and it was also implied (although not strictly modeled)
for SN 2017cbv \citet{griffin}.

\subsection{Model Grids}\label{ssec:model_grids}

We also compare our lightcurves to two different model grids:
the double detonation models of \citet{polin_double_det}
and the Ni distribution models of \citet{magee_models}.
We use the same epoch range as in Section \ref{ssec:blue_bumps},
i.e. all data earlier than $t_{\rm{max}}+5$ days.
This tests if the models can reproduce the rise time and peak magnitudes across multiple filters,
while still weighting the earlier data more than if we included the full lightcurve.

\subsubsection{Double detonation}\label{sssec:double_det_models}

In the \citet{polin_double_det} models,
the explosion mechanism is as follows:
a layer of He is built up on the surface of a sub-Chandrasekhar mass WD (mass range 0.6--1.3 $\Msun$).
This shell is then detonated at a range of masses (0.01--0.1 $\Msun$).
As the He shell detonates this drives a shock into the WD which causes it to detonate.
This can produce SNe~Ia at a range of absolute magnitudes,
from dimmer than $-15$ to brighter than $-19$,
and also with a range of potentially bumpy behavior at early times.

The model grid consists of 43 models, with the parameter ranges described above.
For each SN in our sample we correct the data to the distance modulus, extinction, and time of first light
detailed in Sections \ref{sec:analysis} and \ref{sssec:rise_times}.
We compare the lightcurve of each object to each model across all available filters ($UBVgri$),
using a simple reduced $\chi^2$ metric to score each comparison.
Best-fit models are listed in Table \ref{table:model_grid_best_fits}.

As seen in the table, many of our objects converge to a narrow range of best-fit parameters,
(the most massive WDs with the least massive He shell,
which are representative of the most ``normal" models in the grid).
Indeed multiple objects converge to the exact same model.
We attribute this partially to the coarseness of the model grid:
although one benefit of this model grid is that it provides a uniform scenario which can generate a wide range of SNe~Ia behavior with only two parameters
(WD mass and He shell mass),
a downside is that in the parameter space describing normal SNe~Ia it is not finely sampled enough to discriminate between objects in this sample.

\begin{figure}[t!]
\begin{center}
\includegraphics[width=0.47\textwidth]{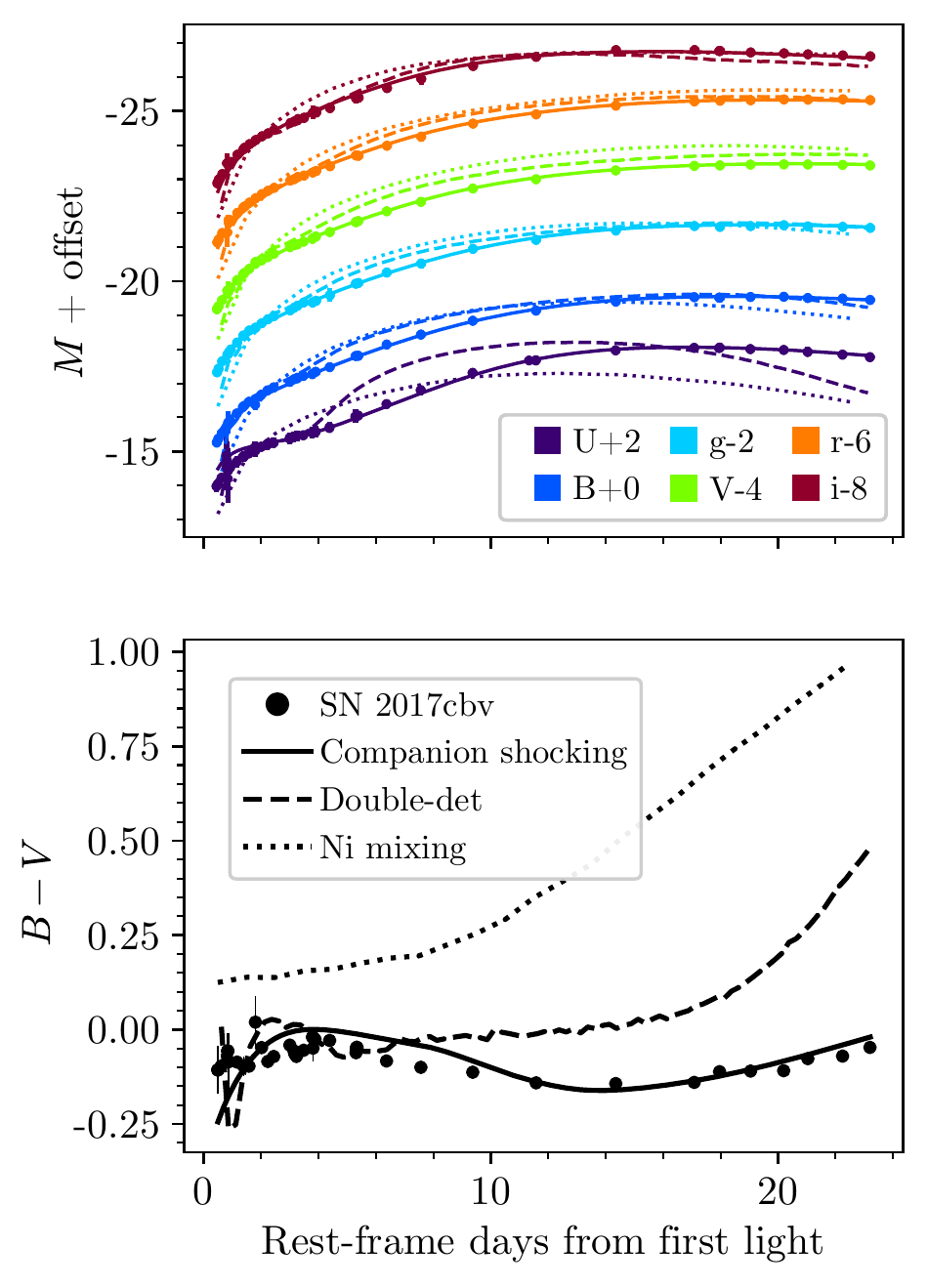}
\caption{
LCO data of SN~2017cbv compared with the best-fit models described in
Sections \ref{ssec:blue_bumps} and \ref{ssec:model_grids}
and listed in
Tables \ref{table:kasen_parameters} and \ref{table:model_grid_best_fits}.
All fits were done in magnitude space (top panel),
but we show them in color space ($B-V$, bottom panel) as well.
The companion shocking models fit better in both cases,
although the model's earliest $B-V$ colors evolve more than they are observed to do 
(probably due to oversimplified temperature evolution, as discussed in the text).
}
\label{fig:17cbv_model_comp}
\end{center}
\end{figure}

The best-fit double-detonation model for SN~2017cbv is plotted alongside its data in Figure \ref{fig:17cbv_model_comp} as the dashed line.
This object is again chosen because it displays the strongest early excess in the sample,
and thus makes a good test for the model grids' flexibility.
In magnitude space,
the model does indeed have a bump of similar strength and duration to the one seen in SN~2017cbv,
and it provides a close match to $i$-band data throughout.
The main discrepancy is that the bluer bands rise too quickly after the early excess,
and extending the model out to later times shows a decline in the bluer bands which is much faster than observed in the data.
This behavior is generally representative across the sample,
with the bluer filters (especially \textit{U}) of best-fit models peaking earlier than the data by multiple days.
This could partially be alleviated by introducing a $\Delta t_{U}$ parameter as we did for the companion shocking models,
but the average shift for the sample there was 0.33 days,
a small recalibration compared to the systematic shift of multiple days required for these models.
The fit was done in magnitude space,
but looking at the color evolution of the model in Figure \ref{fig:17cbv_model_comp},
the color during the excess is closely matched 
(although with more evolution than is seen in the data),
but past the excess the model is systematically redder than the data,
with the discrepancy increasing over time.

\begin{deluxetable}{cccc}[t!]
\tablehead{
\colhead{SN} & \colhead{Double-det} & \colhead{Ni mixing} &
}
\startdata
2017cbv & 1.1, 0.05 & \texttt{EXP\_Ni0.8\_KE2.18\_P3} \\
2017cyy & 1.1, 0.01 & \texttt{DPL\_Ni0.8\_KE1.81\_P3} \\
2017erp & 1.2, 0.01 & \texttt{DPL\_Ni0.8\_KE1.68\_P3} \\
2017fgc & 1.1, 0.01 & \texttt{DPL\_Ni0.8\_KE0.65\_P3} \\
2018gv  & 1.1, 0.01 & \texttt{DPL\_Ni0.8\_KE1.68\_P3} \\
2018xx  & 1.1, 0.01 & \texttt{DPL\_Ni0.8\_KE1.81\_P3} \\
2018yu  & 1.2, 0.01 & \texttt{DPL\_Ni0.8\_KE1.68\_P3} \\
2018aoz & 1.1, 0.01 & \texttt{DPL\_Ni0.8\_KE1.68\_P3} \\
2019np  & 1.1, 0.01 & \texttt{DPL\_Ni0.8\_KE1.68\_P3} \\
\enddata
\caption{
Best-fit models in the two model grids described in Section \ref{ssec:model_grids}.
For the double-detonation models we report the best-fit models as a tuple of 
WD mass in $\Msun$,
He layer mass in $\Msun$.
We use the same shorthand as \citet{magee_models} for the Ni mixing models:
exponential (\texttt{EXP}) or
double power law (\texttt{DPL}) density profile,
Ni mass in $\Msun$ (0.4 -- 0.8),
KE in foe (0.50 -- 2.18),
and a parameter which determines the degree of mixing,
ranging from 3 (most mixed) to 
100 (least mixed).
}
\label{table:model_grid_best_fits}
\end{deluxetable}

\subsubsection{Ni mixing}\label{sssec:ni_models}

We also look for best-fit models from the grid of 255 models provided by \citet{magee_models}.
These models make use of the radiative transfer code \texttt{TURTLS} \citep{turtls} and vary the 
density profiles, Ni masses, kinetic energy, and degree of Ni mixing
to produce a range of lightcurves up to $+25$ days from the explosion.

Rather than finding the best-fit model with a direct $\chi^2$ search 
we marginalize over parameters following the methodology of \citet{daichi_short_plateau} (see Section 4.3 and Figure 7 of that paper).
As above,
we do the fits in absolute magnitude space using the parameters from 
Sections \ref{sec:analysis} and \ref{sssec:rise_times},
and we report the results in Table \ref{table:model_grid_best_fits}.
There is a similar problem to the double-detonation grid,
i.e. that multiple objects converge to the same fits,
in this case those with the highest Ni mass which is mixed most completely.

Again the best-fit Ni mixing model for SN~2017cbv is plotted alongside the data in Figure \ref{fig:17cbv_model_comp}, as the dotted lines.
As noted in \citet{magee_models},
models in this grid struggle to reproduce any early excess.
A separate paper was published specifically modeling SNe 2017cbv and 2018oh \citep{magee_maguire},
and clumps of Ni (0.02 -- 0.04 $\Msun$) in the outer ejecta were required to replicate the excesses.
Indeed the best-fit model here cannot replicate the early excess,
and misses the data on other counts as well
(e.g. $U$-band peak magnitude and time of maximum are both underpredicted,
$V$-band is systematically overpredicted,
all bands rise too quickly after the epochs of the excess).
In color space
the model is systematically redder than the data,
with the difference increasing to a full magnitude at the latest phase shown in Figure \ref{fig:17cbv_model_comp}.

The fact that this model grid does not contain any early excesses is a detriment for our purposes,
as it makes it difficult to accurately parametrize objects with early excesses.
On the other hand,
it makes it a possible diagnostic of whether or not there is an early excess:
when \citet{miller_rise_times} fit general power laws to the rising lightcurves of a sample of 127 SNe~Ia observed with ZTF,
they found no instances of early excesses.
But \citet{deckers_ztf_excesses},
reexamining 115 SNe~Ia from that sample with the model grid here,
were able to recover six instances of early excesses based on quantitative measures of whether all models produced poor fits to the earliest epochs.
This also shows that the statement ``this SN does/does not have an early excess" can be dependent on the methodology used to characterize these excesses.
We believe this is why so few objects in the literature have reported early excesses:
you both need high-cadence multiwavelength data $\sim$3 magnitudes fainter than peak
(within $\sim$5 days of first light),
and you need a methodology able to detect and characterize these excesses,
except in the extreme cases where they are detectable by eye
as they were for e.g. iPTF14atg, SN 2017cbv, SN 2019yvq, and SN 2021aefx as discussed in Section \ref{sec:intro}.

\section{Color Evolution}\label{sec:color}

\begin{figure*}[t!]
\begin{center}
\includegraphics[width=\textwidth]{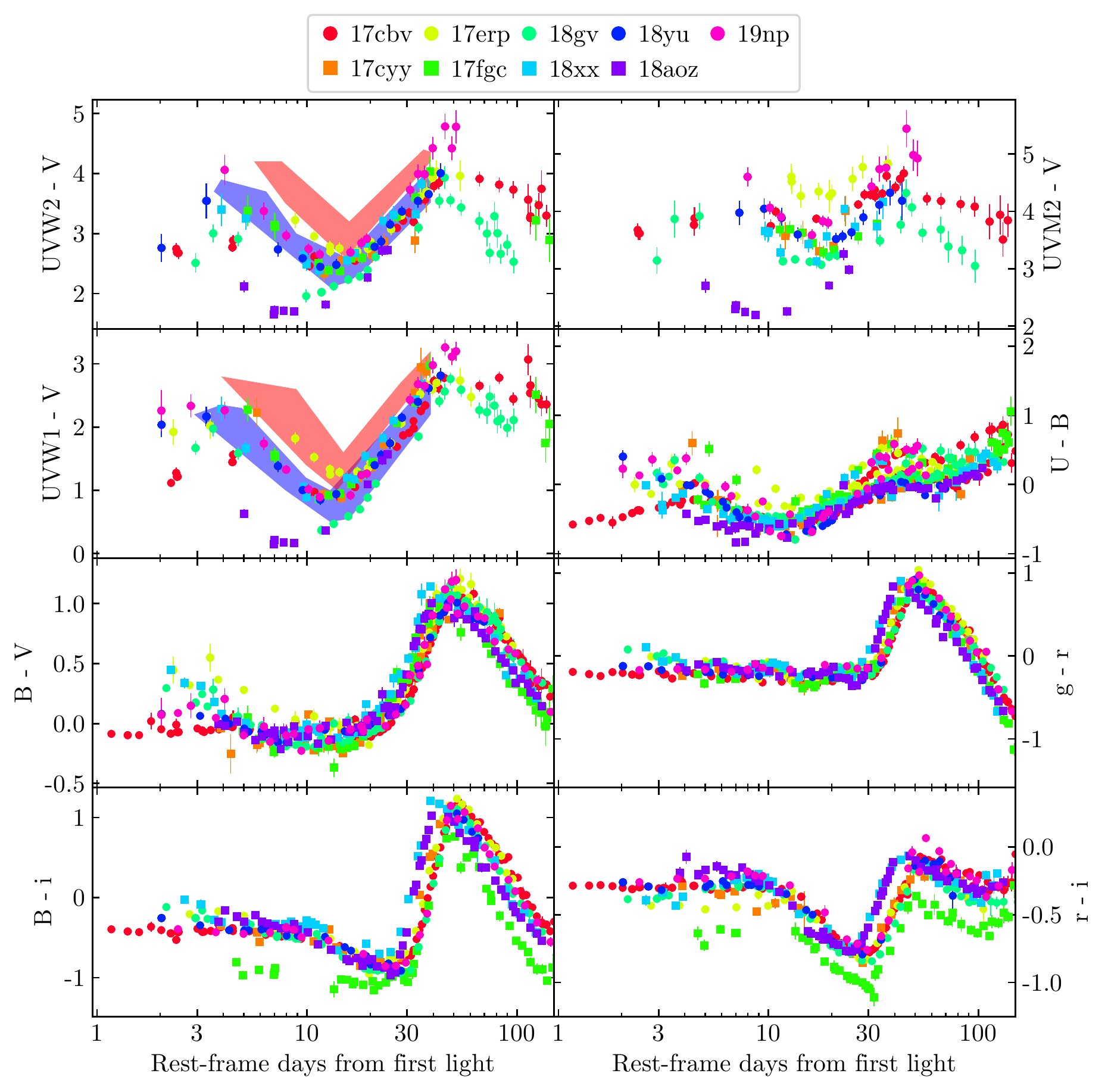}
\caption{
Color evolution of the sample in a number of filter combinations.
Note the logarithmic scale of the time axis,
to emphasize early heterogeneity while showing evolution into the nebular phase.
NUV-red and -blue regions are shown for UVW2--V and UVW1--V \citep[see Figure 4 of][]{milne_uv_colors},
where it can be seen most objects in this sample are NUV-blue,
in contrast with expectations from \citet{milne_uv_colors}.
}
\label{fig:color_evolution}
\end{center}
\end{figure*}

With the extinction values measured in Section \ref{ssec:snoopy},
we present the color evolution of the sample in Figure \ref{fig:color_evolution}.
We use a logarithmic time axis (from one day after first light)
to distinguish any early inhomogeneities.
We also include the NUV-red and -blue regions defined in \citet{milne_uv_colors} for the appropriate filter combinations.

As can be seen in the figure,
the objects in the sample are all NUV-blue with the possible exception of SN~2017erp before peak brightness \citep[see][which also found that the object was both reddened and intrinsically red]{brown_2017erp}.
This is unexpected:
\citet{milne_uv_colors} found that only about a third of their sample were NUV-blue.
Even calling SN~2017erp NUV-red would mean that 8 of 9 objects in this sample are NUV-blue.
The probability of this happening (according to the binomial distribution) is only 0.08\%.
\citet{brown_uv_colors} noted that the exact way reddening is applied can easily shift objects between NUV color groups,
and even a small amount of optical reddening ($E(B-V)=0.2$)
can drastically change the UV colors of an object.
Our sample,
which is largely consistent with $E(B-V)=0$ (see Table \ref{table:lc_params}),
would only get bluer if reddening were underestimated (and it could not be overestimated).
This could indicate that the sample of \citet{milne_uv_colors} were incorrectly dereddened,
and that SNe~Ia are actually UV-bluer than previously thought.
Additionally, not only are the objects not in the NUV-red group,
but a significant portion of the data earlier than $+10$ days from first light is bluer than the bluest edge of the expected NUV-blue population:
again, SNe~Ia seem UV-bluer than previously thought.

In addition to the already-known $>$1 magnitude scatter in the NUV colors,
we also note the range of $B-V$ colors in the first $\sim$5 days after first light.
Here the object with the earliest data (SN~2017cbv) has a steady mostly unchanging blue color close to $B-V=-0.1$.
This contrasts with the color evolution of the object with the next-earliest data (SN~2018yu),
which at its earliest phase ($+2.0$ days) has a redder color ($B-V=0.08 \pm 0.03)$.
Another object with especially good early data (SN~2018xx, $+2.3$ days) has significantly redder colors ($B-V=0.45 \pm 0.03$).

\begin{figure}[t]
\begin{center}
\includegraphics[width=0.47\textwidth]{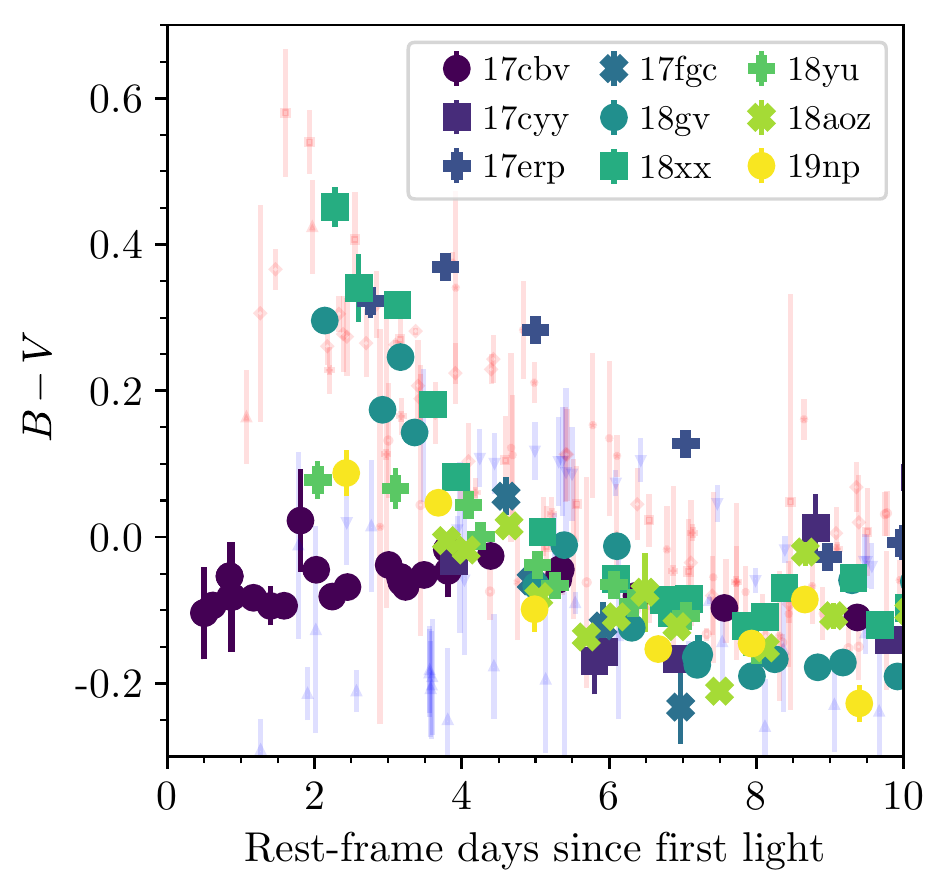}
\includegraphics[width=0.47\textwidth]{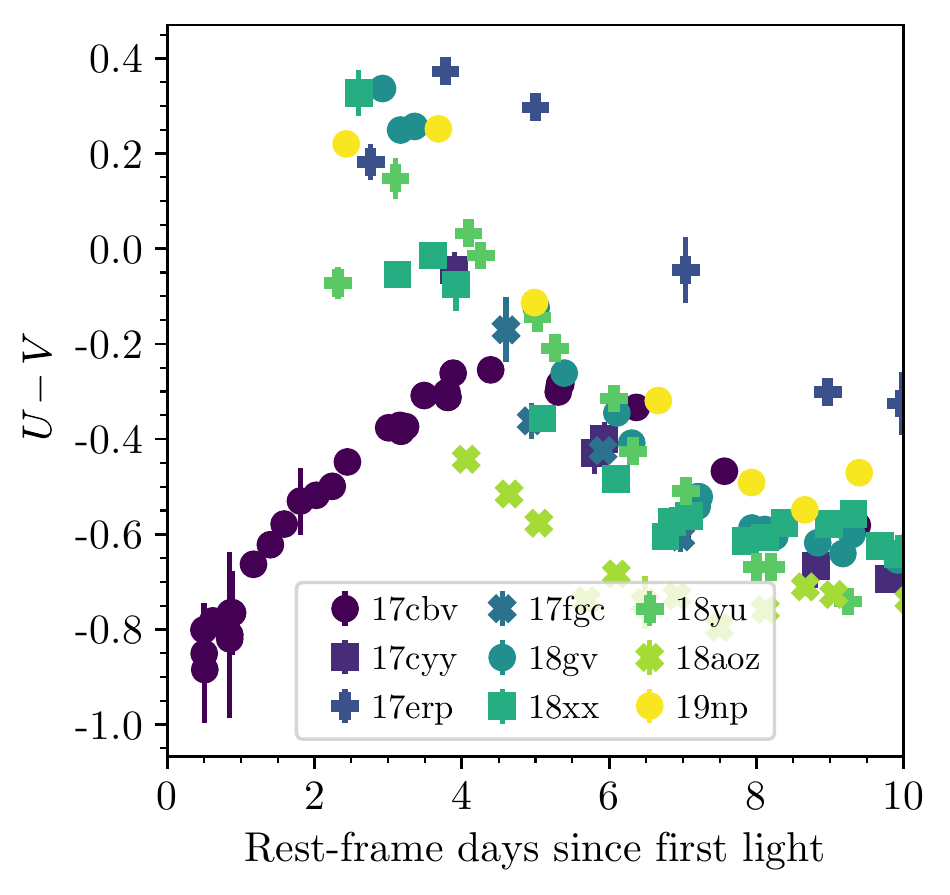}
\caption{
\textit{Top:} 
Early $B-V$ color evolution of objects in the sample,
with background red and blue data points from the objects in \citet{redvsblue},
colored according to their classification in Table 1 of that paper.
\textit{Bottom:} 
Early $U-V$ color evolution.
Whereas most objects just show a blueward color evolution,
the three objects with early excesses (SNe 2017cbv, 2017erp, and 2018yu)
have an initial redward evolution from the rapid cooling of the shocked ejecta before becoming dominated by the usual blueward evolution.
}
\label{fig:redvsblue}
\end{center}
\end{figure}

The top panel of Figure \ref{fig:redvsblue} shows the early $B-V$ color evolution
compared to the objects in \citet{redvsblue},
since that paper includes several other SNe~Ia with early multiband data.
\citet{redvsblue} propose two distinct populations of SNe~Ia given their color evolutions within $\sim$4 days of explosion,
and we have kept this red/blue dichotomy when coloring the background points of the figure (to replicate their Figure 2).
We choose a neutral colormap for our sample to avoid confirmation bias in seeing two distinct populations especially in the critical phase (earlier than 4 days after first light).

We find a range of early color behaviors,
with SNe 2018yu and 2019np fitting in between the ``early-blue" objects 
(such as SN~2017cbv)
and the ``early-red" objects,
which SN~2018xx matches with.
Our small sample size
makes it difficult for us to say with any statistical confidence whether there is a real continuum of behavior at the earliest phases,
although
\citet{ztf_colors} find no evidence for multiple distinct populations when examining the early $g-r$ color evolutions of a sample of 65 SNe~Ia.
(We refer especially to Figure 2 of that paper for comparison to the \citet{redvsblue} sample,
although the putative discrepant populations are more distinct in $B-V$ than in $g-r$.)

Figure \ref{fig:redvsblue} also shows the $U-V$ color evolution.
We have kept the same neutral colormap to avoid confirmation bias,
but most objects (e.g. SNe 2018gv, 2018aoz)
show a direct blueward color evolution.
SN 2017cbv shows a very clear and well-resolved initial redward color evolution before overlapping with the blueward evolution after $\approx$5 days.
The other two objects with early excesses (SNe 2017erp and 2018yu) display similar redward turns from between their first and second epochs,
as does SN 2019np \citep[discussed above and in][although again we don't confidently claim it as having an early excess]{2019np_sai}.
These ``kinks" in the color evolution are more obvious in $U-V$ than the range of early slopes in $B-V$,
and presumably arise from the rapid cooling of the shocked ejecta in the early shock-dominated phase,
before matching more typical color evolution in the later SN-flux--dominated regime.

We have already compared the color evolution of objects in our sample with our best-fit models in Figure \ref{fig:17cbv_model_comp} and the surrounding discussion.
As stated there, we favor companion shocking models in color space in addition to magnitude space.

% don't think I need any of this any more? keeping in comments in case
%but many models have a range of early colors for a variety of reasons.
%We list a few examples here.
%Models with varying distributions of Ni in the ejecta can have a range of early colors
%(due to gamma rays from Ni decay being reprocessed to different degrees depending on where they originate),
%see for instance Figure 8 of \citet{piromorozova} or Figure 3 of \citet{magee_maguire}.
%Double-detonation models also have a range of early behaviors depending on the thickness of the He envelope
%(due again to differences in gamma ray reprocessing as the outer He shell burns),
%see e.g. Figure 6 of \citet{polin_double_det}.
%Single-degenerate progenitor systems can also have widely varying early colors depending on the parameters of the system
%(since companions of different sizes, densities, and orbital separation will shock the ejecta to varying degrees),
%see Figure 3 of \citet{kasen}.
%Lastly (in this non-exhaustive list),
%the ejecta can still get shock-heated even in the absence of a nearby stellar companion if there is sufficiently dense circumstellar material,
%again leading to a variety of early colors \citep[see again][Figures 12 and 15]{piromorozova}.
%While we can exclude some regions of parameter space,
%for instance some models in \citet{polin_double_det} have dramatic red bumps at early times not observed in our data,
%beyond that it is difficult to say with certainty which models are preferred based on the color evolution alone.

\section{Discussion}\label{sec:discussion}

As shown in the previous two sections,
SNe~Ia exhibit a continuum of behavior at early times.
This is consistent with the single-degenerate scenario described in \citet{kasen},
where this continuum is caused partly by seeing the progenitor system and its briefly shocked ejecta at a range of viewing angles.

In general we favor the companion interaction scenario for several reasons:
\begin{enumerate}
\item
With effectively only two physical parameters (companion separation and viewing angle)
these models can reproduce a wide range of early lightcurve behavior seen in SNe~Ia.
\item
The number of early excesses and the distribution of their strengths
(one SN with a strong excess, two SNe with weaker excesses, 6 SNe with no early excess)
match their expected viewing-angle-dependent rate
(see Figures \ref{fig:kasen_a_theta_posteriors} and \ref{fig:theta_CDF}).
\item
Referring to Figure \ref{fig:kasen_a_theta_posteriors},
for the three objects in our sample with early bumps
the inferred companion separation lies close to the expected range from binary stellar evolution models.
\item
The prediction that early spectra should appear to be diluted with a blackbody component is matched by observations
\citetext{\citealp[e.g. SN~2012cg in][]{marion},
\citealp[SN~2019yvq in][]{burke_19yvq},
\citealp[and implied for SN~2017cbv in][]{griffin}}.
\item
Although not without systematic issues,
companion interaction models reproduce both the magnitude and color evolution better than other classes of models 
(see Figures \ref{fig:17cbv_model_comp} and \ref{fig:kasen_residuals}).
\item
Lastly (and more philosophically) the models are true theoretical predictions,
as this method of producing early excesses was formulated before the effect had ever been observed,
meaning best-fit parameters lie close to a priori expected values without any fine-tuning needed to match observations.
\end{enumerate}

As with any models there are of course imperfections:
\begin{enumerate}
\item
These models are an analytic approximation to the full numerical grid of \citet{kasen}
\citep[including the semi-analytic viewing angle effect from][]{brown_2012},
and this results in certain oversimplifications such as an inaccurate temperature evolution as discussed in Section \ref{sssec:kasen_best_fits}.
The models also have a grey opacity,
which means the SN ejecta do not reprocess the UV shock flux for off-axis explosions,
leading to dramatic overestimates of the flux in Swift filters for multiple SNe \citep{griffin,griffin_2021aefx}.
\item
The earliest and most interesting phase of the models relies on comparing to the $s=1$ template at precisely the phase when it is most poorly constrained.
%and these templates are difficult to accurately produce in the near-UV 
%(where the most interesting effects are).
\item
The fits for objects without early excesses
(see Table \ref{table:kasen_parameters} and Figure \ref{fig:kasen_a_theta_posteriors})
tend to converge to perfectly misaligned systems,
since the distinction between e.g. $\theta=135\degree$ and $\theta=180\degree$ is lost in the noise (see Figure \ref{fig:theta_function}).
This limits the range of $\theta$ and thus the sample size where we can compare to the expected viewing angle distribution.
\item
As discussed in Section \ref{sec:intro},
a nondegenerate companion is expected to leave traces of H in the nebular spectra of SNe~Ia which originate from single-degenerate systems, 
and this signature is not detected in $>$100 SNe~Ia 
\citep[e.g.][]{dave_nebular_halpha,tucker_nebular_halpha}.
This signature has only ever been observed in three low-luminosity SNe~Ia,
and notably has \textit{not} been observed for objects with early excesses such as
SN~2017cbv \citep{dave_17cbv_nebular},
SN~2018oh \citep{dimitriadis_18oh_nebular,tucker_18oh_nebular},
SN~2021aefx \citep{ashall_2021aefx,griffin_2021aefx},
or for any objects in this sample (Sand et al. 2022, in prep.).
However,
measuring a physical amount of stripped H ($M_{\rm{H}}$) relies on translating from a measured luminosity of H$\alpha$ ($L_{\rm{H}\alpha}$),
and this translation is extremely model-dependent:
early models \citep{mattila_nebular_H}
calculate a luminosity for a given $M_{\rm{H}}$ which is up to $10^{3}$ times less than 
later models \citep{botyanski_halpha} calculated for the same $M_{\rm{H}}$.
Even more recent models \citep{dessart_nebular_H} find variations of up to half a dex
in $L_{\rm{H}\alpha}$
depending on the adopted model of the underlying SN ejecta,
and the time-dependent nature of $L_{\rm{H}\alpha}$ can again cause the inferred $M_{\rm{H}}$ to differ by a full order of magnitude from the \citet{botyanski_halpha} models
(see Sand et al. 2022, in prep.).
The models of \citet{botyanski_halpha} are three-dimensional and do not assume local thermodynamic equilibrium (LTE),
but make several approximations to optimize the radiative transfer calculations
including using the Sobolev approximation and limiting the allowed recombination transitions;
while \citet{dessart_nebular_H} has more complete radiative transfer calculations,
those models are limited to one dimension,
with an ``unphysical" treatment of how the stripped H is added.
All this is to say:
while the lack of H$\alpha$ observed in the nebular spectra of most SNe~Ia is seemingly in tension with our results that early excesses arise from companion interaction,
the extreme range of highly model-dependent $L_{\rm{H}\alpha}$ predictions lessens this tension and we certainly do not take the lack of H$\alpha$ as a reason to definitively reject the companion interaction scenario.
\end{enumerate}

Regardless of these downsides,
in the interest of having a consistent methodology for fitting the rising lightcurves of all SNe~Ia regardless of whether they show signs of interaction,
we have fit the objects in our sample with these companion interaction models.
We generally favor them over other classes of models,
especially in cases where objects show early excesses,
and we find that our sample is holistically consistent with Roche-lobe-overflowing single-degenerate progenitor systems described by companion interaction models.

\section{Conclusions} \label{sec:conclusions}

We have presented a sample of nine SNe~Ia with exemplary early-time high-cadence multiwavelength followup from LCO and Swift.
We have relied on other facilities for SN discovery,
and the DLT40 survey with its nearby-galaxy-targeted approach enabled much of the earliest and most interesting observations presented here.

Out of the nine objects in the sample,
one has a strong early excess (SN~2017cbv),
two have weaker excesses (SNe 2017erp and 2018yu),
and six show no excess at all.
All objects are well-modeled by companion interaction models,
which add an early viewing-angle-dependent shock component to a standard SN Ia template.
Even for cases where these models are less successful at fitting the data
\citep[i.e. SN 2019np and the earliest data of SN 2018aoz, see][]{2019np_sai,ni_18aoz},
the discrepancies reveal an exciting diversity of early lightcurve behavior.
The sample size is small,
but the strengths of early excesses and the distribution of viewing angles are statistically consistent with SNe~Ia predominantly arising from Roche-lobe-overflowing single-degenerate progenitor systems.

In addition to the companion interaction results,
we also find that eight of the nine SNe are near-UV blue,
in contrast to the expectation that only a third of SNe~Ia are near-UV blue.
We also find a seemingly continuous range of $B-V$ colors within the first five days from first light,
again in contrast to earlier claims of two distinct populations based on early color evolution.

The data required to reveal unusual early lightcurve behavior and distinguish between models are extremely difficult to get:
the data must be early, high cadence, and preferably multiwavelength.
LCO is uniquely suited to obtain just this kind of dataset -- no other ground-based optical observatory could have obtained the continuous 6-hr cadence coverage of SN~2017cbv's bump,
and its rapid response allowed the most interesting data of SNe 2021aefx, 2018yu, 2017erp, and the other objects here.
The uncertainties surrounding the progenitor systems of SNe~Ia will slowly decrease over time
as samples such as the one presented here,
with early high-cadence multiwavelength data,
gradually expand,
allowing for testing of the finer details and statistical predictions of a variety of models.

\acknowledgments

We are grateful to A. A. Miller for making the code used in \citet{miller_rise_times} open source,
allowing us to test its use for fitting the rising lightcurves of our sample
(even though we ultimately did not adopt this methodology).
We are also grateful to him for providing the requested MCMC posteriors when we wished to compare to results from our sample.

J.B., D.A.H., C.M., and C.P. are supported by NSF grants AST-1911151 and AST-1911225, as well as by NASA grant 80NSSC19kf1639.
L.T. acknowledges support from MIUR (PRIN 2017 grant 20179ZF5KS).

Time domain research by the University of Arizona team and D.J.S.\ is supported by NSF grants AST-1821987, 1813466, 1908972, \& 2108032, and by the Heising-Simons Foundation under grant \#2020-1864.

Research by Y.D., and S.V.\ is supported by NSF grants AST-1813176 and AST-2008108.

P.J.B. is supported by NASA grants 80NSSC-20K0456 and 80NSSC19K0316.

K.A.B acknowledges support from the DIRAC Institute in the Department of Astronomy at the University of Washington. The DIRAC Institute is supported through generous gifts from the Charles and Lisa Simonyi Fund for Arts and Sciences, and the Washington Research Foundation

This research makes use of the NASA/IPAC Extragalactic Database (NED) which is operated by the Jet Propulsion Laboratory, California Institute of Technology, under contract with NASA.
It also makes use of observations from the Las Cumbres Observatory network.

\vspace{5mm}
\facilities{Las Cumbres Observatory (Sinistro), Swift (UVOT)}

\software{
\texttt{astropy } \citep{2013A&A...558A..33A,astropy},
\texttt{SNooPy} \citep{snoopy},
\texttt{lightcurve\_fitting} \citep{griffin_lightcurvefitting},
\texttt{emcee} \citep{emcee},
\texttt{extinction} \citep{extinction_zenodo}
          }

\bibliographystyle{aasjournal}
\bibliography{earlyIa}

\end{document}